\documentclass[11pt,]{article}

\usepackage{PRIMEarxiv}
\usepackage[affil-it]{authblk}
\usepackage[utf8]{inputenc} 
\usepackage[T1]{fontenc}    
\usepackage{hyperref}       
\usepackage{url}            
\usepackage{booktabs}       
\usepackage{amsfonts}       
\usepackage{nicefrac}       
\usepackage{microtype}      
\usepackage{lipsum}
\usepackage{graphicx}
\graphicspath{{media/}}     

\newcommand{\Figure}[1]{Figure~\ref{fig:#1}}

\newcommand{\Table}[1]{Table~\ref{tab:#1}}

\newcommand{\Section}[1]{Section~\ref{sec:#1}}

\usepackage{amsmath}
\usepackage{bbold}
\renewcommand\vec{\mathbf}
\newcommand{\mat}{\mathbf}

\DeclareMathOperator*{\argmin}{argmin}

\usepackage{enumitem}

\usepackage{caption}
\usepackage{subcaption}

\usepackage{multirow}
\usepackage{graphicx}

\usepackage{sidecap}

\usepackage{float}

\usepackage[flushleft]{threeparttable}

\usepackage{amssymb}

\usepackage[linesnumbered,ruled,vlined]{algorithm2e}
\usepackage {algpseudocode}
\usepackage{algorithmicx}
\usepackage{algcompatible}

\newcommand{\nosemic}{\renewcommand{\@endalgocfline}{\relax}}
\newcommand{\dosemic}{\renewcommand{\@endalgocfline}{\algocf@endline}}
\let\oldnl\nl
\newcommand{\nonl}{\renewcommand{\nl}{\let\nl\oldnl}}

\author[1,2]{\textbf{George Yiasemis}}
\author[1,2]{\textbf{Nikita Moriakov}}
\author[1]{\textbf{Jan-Jakob Sonke}}
\author[1,2,3]{\textbf{Jonas Teuwen}}

\affil[1]{Department of Radiation Oncology, the Netherlands Cancer Institute,  Amsterdam, the Netherlands}
\affil[2]{University of Amsterdam, Amsterdam, the Netherlands}
\affil[3]{Radboud University Medical Center, Department of Medical Imaging, Nijmegen, the Netherlands}

\title{vSHARP: variable Splitting Half-quadratic Admm algorithm for Reconstruction of inverse-Problems}

\begin{document}
\maketitle

\begin{abstract}
    Medical Imaging (MI) tasks, such as accelerated parallel Magnetic Resonance Imaging (MRI), often involve reconstructing an image from noisy or incomplete measurements. This amounts to solving ill-posed inverse problems, where a satisfactory closed-form analytical solution is not available. Traditional methods such as Compressed Sensing (CS) in MRI reconstruction can be time-consuming or prone to obtaining low-fidelity images. Recently, a plethora of Deep Learning (DL) approaches have demonstrated superior performance in inverse-problem solving, surpassing conventional methods. In this study, we propose vSHARP (variable Splitting Half-quadratic ADMM algorithm for Reconstruction of inverse Problems), a novel DL-based method for solving ill-posed inverse problems arising in MI. vSHARP utilizes the Half-Quadratic Variable Splitting method and employs the Alternating Direction Method of Multipliers (ADMM) to unroll the optimization process. For data consistency, vSHARP unrolls a differentiable gradient descent process in the image domain, while a DL-based denoiser, such as a U-Net architecture, is applied to enhance image quality. vSHARP also employs a dilated-convolution DL-based model to predict the Lagrange multipliers for the ADMM initialization. We evaluate vSHARP on tasks of accelerated parallel MRI Reconstruction using two distinct datasets and on accelerated parallel dynamic MRI Reconstruction using another dataset. Our comparative analysis with state-of-the-art methods demonstrates the superior performance of vSHARP in these applications.
\end{abstract}

\keywords{Medical Imaging Reconstruction \and Inverse Problems \and Deep MRI Reconstruction \and Mathematical Optimization \and Half-Quadratic Variable Splitting \and Alternating direction method of multipliers}

\section{Introduction}
\label{sect:intro}
Medical Imaging (MI) possesses a wide range of modalities that can by employed in the clinic that can assist in diagnosis, treatment, and treatment planning of patients. The need to acquire high-quality images in a timely manner has been a driving force behind many medical advancements. Inverse problems play a pivotal role in this pursuit, providing a framework to reconstruct accurate images from the limited and often noisy measurements obtained through imaging modalities such as  Magnetic Resonance Imaging (MRI). These inverse problems have been the focal point of intensive research, as they lie at the intersection of mathematical modeling, signal processing, and MI.

Traditionally, solving inverse problems arising in MI has relied on well-established mathematical techniques and iterative numerical algorithms. These methods involve imposing prior knowledge about the imaging process, such as sparsity or smoothness constraints, to regularize the ill-posed nature of the problem. In the context of accelerated parallel MRI reconstruction, for instance, techniques like Compressed Sensing (CS) have been employed to recover high-resolution images from undersampled $k$-space data \cite{Glockner2005}. 

Recent years have witnessed a transformative shift in the field of medical imaging, largely driven by advancements in the realms of Deep Learning (DL) and Computer Vision. DL-based techniques, particularly convolutional neural networks (CNNs), have shown remarkable potential in solving a diverse range of inverse problems \cite{ongie2020deep,raissi2019physics,8253590}. These methods leverage the capacity of CNNs to learn complex relationships directly from data, eliminating the need for hand-crafted mathematical models and heuristics. This shift has enabled the development of innovative solutions that surpass the limitations of traditional methods in MI. By leveraging large datasets, DL-based reconstruction methods can solve inverse problems by learning complex image representations and effectively reconstruct high-fidelity images, often in a supervised learning (SL) settings \cite{8434321,Yiasemis_2022_CVPR,10.1007/978-3-030-59713-9_7, SOUZA2020140,moriakov2022lire,zhang2021improving,madesta2020self} or in self-supervised settings \cite{Hendriksen_2020,lehtinen2018noise2noise}. 

In this study, we propose vSHARP (variable Splitting Half-quadratic ADMM algorithm for Reconstruction of inverse-Problems), an innovative approach to solving inverse problems in Medical Imaging using Deep Learning. The vSHARP method leverages the Half-Quadratic Variable Splitting (HQVS) \cite{CHENG2020103193} method and integrates the Alternating Direction Method of Multipliers (ADMM) \cite{CAI201821} to unroll an optimization process.  It incorporates a differentiable gradient descent process for data consistency in the image domain, a denoising process using deep learning-based denoisers such as the U-Net architecture \cite{ronneberger2015unet} for image enhancement, and a trained initializer for the ADMM Lagrange multipliers. The goal of vSHARP is to achieve high-quality reconstructions.

To evaluate the performance of vSHARP, we conducted comprehensive experiments focused on accelerated parallel MRI reconstruction, applying vSHARP to the Calgary Campinas brain dataset \cite{Beauferris2022} and the fastMRI T2 prostate dataset \cite{tibrewala2023fastmri}. Additionally, we tested accelerated parallel dynamic MRI reconstruction on the CMRxRecon cardiac cine dataset \cite{cmrxrecondataset}.

Our main contributions can be summarized as follows:

\begin{itemize}
    \item We propose the vSHARP algorithm, a novel versatile DL-based inverse problem solver that can be applied on inverse problems arising in (Medical) Imaging with well-defined forward and adjoint operators.
    \item We provide a comprehensive mathematical derivation of vSHARP using the general formulation of inverse problems and employing the HQVS and ADMM methods.
    \item We assess vSHARP's performance in the context of accelerated parallel static and dynamic MRI reconstruction, utilizing brain, prostate and cardiac MRI datasets alongside different sub-sampling schemes. Our results demonstrate vSHARP's exceptional capabilities, surpassing current baseline methods.
\end{itemize}

\section{Theoretical Background}
\label{sec:sec2}

In imaging, the concept of an inverse problem involves the task of recovering an underlying image $\vec{x} \in \mathcal{X}$ from given noisy or incomplete measurements $\tilde{\vec{y}} \in \mathcal{Y}$, where $\mathcal{X}, \mathcal{Y}$ are Hilbert spaces, with $\mathcal{X}$ denoting the space of functions representing possible images and $\mathcal{Y}$ the space of functions representing measurements. These measurements, obtained through a known process known as the forward model, are related according to the equation
\begin{equation}
    \tilde{\vec{y}} = \mathcal{A} \left( \vec{x} \right) + \boldsymbol{\epsilon},
    \label{eq:forward_model}
\end{equation}
where $\mathcal{A}:\mathcal{X} \rightarrow \mathcal{Y}$ denotes the forward measurement operator. Here we assume that $\boldsymbol{\epsilon} \sim \mathcal{N}(\boldsymbol{0}, \mathbf{I}_{\mathcal{Y}})$ represents additive measurement noise, although this might not be the case for some inverse problems. The operator $\mathcal{A}$, known as the forward operator, is often a composition of various processes, such as blurring, undersampling, sensor sensitivity, and others, depending on the specific problem. 

The adjoint \cite{Conway2007} operator of $\mathcal{A}$ is denoted as $\mathcal{A}^{*}:\mathcal{Y} \rightarrow \mathcal{X}$ and is defined as the unique linear operator such that
\begin{equation}
    \left<\mathcal{A} \left( \vec{x} \right), \, \vec{y}\right>_{\mathcal{Y}} \, = \, \left<  \vec{x} , \, \mathcal{A}^{*} \left( \vec{y} \right)\right>_{\mathcal{X}},\quad \forall \vec{x} \in \mathcal{X},\, \vec{y} \in \mathcal{Y},
    \label{eq:adjoint_formulation}
\end{equation}
where $\left< \cdot, \cdot \right>_{\mathcal X}, \left< \cdot, \cdot \right>_{\mathcal Y}$ are inner products on $\mathcal X, \mathcal Y$ respectively. The adjoint operator is easy to evaluate for modalities such as Magnetic Resonance Imaging and Computed Tomography, and it allows us to reconstruct an image from observed measurements as $\mathcal{A}^{*} \left( \tilde{\vec{y}} \right)$. However, $\mathcal{A}^{*}$ is not necessarily equal to the inverse $\mathcal{A}^{-1}$ of $\mathcal{A}$, even when $\mathcal{A}$ is invertible. In general, due to the noise $\boldsymbol{\epsilon}$ introduced by the measurement process and the fact that $\mathcal{A}$ might have nontrivial kernel (e.g. in accelerated MRI or sparse-view CT), the problem stated in \eqref{eq:forward_model} becomes ill-posed \cite{Kabanikhin2011}. 

Instead of trying derive an analytical approximation to $\mathcal{A}^{-1}$ to arrive at an approximate reconstruction of $\vec{x}$, a common approach is to formulate and solve a regularized least squares optimization problem, combining data fidelity and regularization terms:
\begin{equation}
    \vec{x}^{*} = \argmin_{\vec{x} \in \mathcal{X}} \left| \left| \mathcal{A}(\vec{x}) - \vec{y} \right| \right|_2^2 + \lambda  \mathcal{R}(\vec{x}).
\label{eq:inverse_problem}
\end{equation}
Here,  the regularization functional $\mathcal{R}:\mathcal{X} \rightarrow \mathbb{R}$ encodes prior knowledge about the underlying image, and $\lambda > 0$ controls the trade-off between data fidelity and regularization strength. Solving this optimization problem entails employing numerical techniques, often iterative algorithms that alternate between forward and adjoint operations, aiming to converge towards a solution that effectively balances measurement fidelity and adherence to prior image information.

Common optimization methods for solving \eqref{eq:inverse_problem} include gradient-based techniques like Gradient Descent, Proximal Gradient \cite{doi:10.1137/080716542}, and Conjugate Gradient \cite{Shuai2020}. More sophisticated algorithms such as ADMM \cite{8186925} and Primal-Dual hybrid gradient \cite{Chambolle2010} are also utilized. In this work, we propose an approach that combines the Half-Quadratic Variable Splitting  algorithm with an Alternating Direction Method of Multipliers formulation.

\subsection{Variable Half-quadratic Splitting}

We employ the HQVS method \cite{CHENG2020103193} by first introducing an auxiliary variable $\vec{z} \in \mathcal{X}$ and reformulating \eqref{eq:inverse_problem} in the following equivalent form:
\begin{equation}
    \vec{x}^{*} = \argmin_{\vec{x} \in \, \mathcal{X} } \left( \min_{\vec{z} \in \mathcal{X}, \vec{z} - \vec{x} = 0} \, \frac{1}{2} \big | \big | \mathcal{A} \left(\vec{x}\right) - \tilde{\vec{y}} \big | \big |_2^2 + \lambda\, \mathcal{R}(\vec{z}) \right).
    \label{eq:hqvs_problem}
\end{equation}
This constrained optimization problem can be solved by applying the augmented Lagrangian method \cite{5445028}, i.e.,
\begin{equation}
    \vec{x}^{*} = \argmin_{\vec{x} \in \mathcal{X}} \left( \min_{\vec{z} \in \mathcal{X}} \max_{\vec{u} \in \mathcal{X}}  \mathcal{L}_{\rho}(\vec{x}, \vec{z}, \vec{u}) \right),
    \label{eq:lagrangian_problem}
\end{equation}
where $\mathcal{L}_\rho$ denotes the augmented Lagrangian for \eqref{eq:hqvs_problem}, defined as:
\begin{equation}
\begin{split}
        \mathcal{L}_{\rho}(\vec{x}, \vec{z}, \vec{u}) = \frac{1}{2} \big | \big | \mathcal{A}(\vec{x}) &- \tilde{\vec{y}} \big | \big |_2^2 + \lambda\, \mathcal{R}(\vec{z}) \\
        &+ \left< \vec{u}, \vec{x} - \vec{z} \right>_{\mathbb X} + \frac{\rho}{2} \big | \big | \vec{x} - \vec{z} \big | \big |_2^2,
\end{split}
\end{equation}
where $\vec{u}$ denotes the Lagrange multipliers and $\rho > 0$ is a penalty parameter. 

\subsection{Unrolling via ADMM}

Equation \ref{eq:hqvs_problem} can be iteratively solved over $T$ iterations by applying the ADMM algorithm which comprises the following updates:
\begin{subequations}
    \begin{equation}
        \vec{z}^{t+1}  = \argmin_{\vec{z} \in \mathcal{X}}\, \lambda \, \mathcal{R}(\vec{z}) + \frac{{\rho}_{t+1}}{2} \big | \big | \vec{x}^{t} - \vec{z} + \frac{\vec{u}^t}{{\rho}_{t+1}} \big | \big |_2^2 \quad \quad \text{\textbf{[z-step]}}
        \label{eq:admm_z}
    \end{equation}
    \begin{equation}
    \begin{split}
        \vec{x}^{t+1}  = \argmin_{\vec{x} \in \mathcal{X}}\, &\frac{1}{2} \big | \big | \mathcal{A}(\vec{x}) - \tilde{\vec{y}} \big | \big |_2^2 \\
        +& \frac{{\rho}_{t+1}}{2} \big | \big | \vec{x} - \vec{z}^{t+1} + \frac{\vec{u}^t}{{\rho}_{t+1}} \big | \big |_2^2 \quad \
\text{\textbf{[x-step]}}
        \label{eq:admm_x}
        \end{split}
    \end{equation}
    \begin{equation}
        \vec{u}^{t+1} = \vec{u}^t + {\rho}_{t+1} (\vec{x}^{t+1} - \vec{z}^{t+1}) \quad \quad \text{\textbf{[u-step]}}
        \label{eq:admm_u}
    \end{equation}
\label{eq:admm}
\end{subequations}
\noindent
where $0 \, \le \, t \, \le \, T-1$ denotes the iteration index. In \ref{sec:appendix1} we provide analytically the ADMM algorithm as well as a more detailed derivation of \eqref{eq:admm}. Note that instead of using a single penalty parameter for all iterations, we employ a varying penalty parameter per iteration: $\boldsymbol{\rho} = (\rho_1, \, \rho_2, \cdots, \, \rho_{T}) \in (\mathbb{R}^{+})^{T}$. 

\section{Methods}
\label{sec:sec3}

\subsection{From Unrolled ADMM to vSHARP}
\label{sec:subsec3.1}

In this section, we introduce the variable Splitting Half-quadratic Algorithm for Reconstruction of inverse-Problems (vSHARP) network. vSHARP is a DL-based approach designed to solve \eqref{eq:admm}. 

At iteration $t$, vSHARP addresses the $\vec{z}$-step in \eqref{eq:admm_z} by learning a representation of $\vec{z}^{t+1}$ based on $\vec{z}^{t}$, $\vec{x}^{t}$, and $\vec{u}^{t}$, using a deep neural network denoiser $\mathcal{R}_{{\boldsymbol\theta}_{t}}$ with trainable parameters ${\boldsymbol\theta}_{t}$:
\begin{equation}
    \vec{z}^{t+1}  \, = \,  \mathcal{R}_{{\boldsymbol\theta}_{t + 1}}( \vec{z}^{t},\,\vec{x}^{t},\,\frac{\vec{u}^t}{{\rho}_{t+1}}).
    \label{eq:zt}
\end{equation}
Options for $\mathcal{R}_{{\boldsymbol\theta}_{t}}$ include architectural choices such as U-Nets \cite{ronneberger2015unet}, Uformers \cite{Wang_2022_CVPR}, or other Deep Learning-based (convolutional) structures like ResNet \cite{he2015deep}, DIDN \cite{9025411}, or simple convolution blocks.

Moving on to the $\vec{x}$-step in \eqref{eq:admm_x}, the closed-form solution can be expressed as
\begin{equation}
    \vec{x}^{t+1} \, = \, \big( \mathcal{A}^{*} \circ \mathcal{A} \, + \, {\rho}_{t+1} \mathbf{I}_{n} \big)^{-1} \big( \mathcal{A}^{*} ( \tilde{\vec{y}} ) \, + \, {\rho}_{t+1} \,\vec{z}^{t+1} \, - \, \vec{u}^{t} \big).
    \label{eq:admm_x_inv}
\end{equation}
However, in practice, computing the operator $ \big( \mathcal{A}^{*} \circ \mathcal{A} \, + \, {\rho}_{t+1} \mathbf{I}_{n} \big)^{-1}$ can be computationally intensive or even infeasible. To address this, vSHARP unrolls \eqref{eq:admm_x} in $T_{\vec{x}}$ steps using Data Consistency via Gradient Descent (DCGD), a differentiable Gradient Descent scheme, as outlined by Algorithm \ref{alg:dcgd}, with step sizes $\boldsymbol{\eta} \in (\mathbb{R}^{+})^{T_{\vec{x}}}$.

\begin{algorithm*}
\caption{$\vec{x}$-step: Data Consistency via Gradient Descent (DCGD)}
\label{alg:dcgd}

\textbf{Input:} \newline 
$\rho>0$, $T_x>0$, step sizes $\boldsymbol{\eta}$ ;
\newline 
From current iteration: $\vec{z}^{t+1}$ ; 
\newline 
From previous iteration: $\vec{x}^{t}$, $\vec{u}^{t}$ ; 

Set $\vec{w}^{0} \, = \, \vec{x}^{t}$ ; 

\For{$s\, = \, 0$ \textbf{ \textup{to}} $T_{x} - 1 $}{
    Calculate gradient update: $\vec{\nabla}_{\vec{w}}^{s+1} \, = \, \mathcal{A}^{*} \Big(\mathcal{A} (\vec{w}^{s}) \, - \tilde{\vec{y}}\Big) \, + \, \rho \, \Big(\vec{w}^{s} \, - \, \vec{z}^{t+1} \, + \, \frac{\vec{u}^{t}}{\rho}\Big)$ ;

    Update  $\vec{w}^{s+1}\, = \, \vec{w}^{s} \, - \, \eta_{s+1} \, \vec{\nabla}_{\vec{w}}^{s+1}$ ;
    } 

Set $\vec{x}^{t+1} \, = \, \vec{w}^{T_x}$ ;

\textbf{Output:} $\vec{x}^{t+1}$ ;
\end{algorithm*}

Lastly, the $\vec{u}$-step in Equation \ref{eq:admm} is straightforward as described by \eqref{eq:admm_u}. This step updates the Lagrange multipliers based on the difference between the current estimates of $\vec{x}$ and $\vec{z}$. The parameter ${\rho}_{t+1}$ controls the step size of the update.

\subsection{Initialization of Variables and Parameters of vSHARP}
\label{sec:subsec3.2}
Proper initialization of the image $\vec{x}^{0}$, auxiliary variable $\vec{z}^{0}$, and Lagrange multiplier $\vec{u}^{0}$
can enhance the convergence of the ADMM algorithm and promote better reconstruction outcomes.

\subsubsection{Image and Auxiliary Variable Initialization}
A reasonable initialization for $\vec{x}^{0}$ and $\vec{z}^{0}$ can be produced by using the observed measurements $\tilde{\vec{y}}$, for instance:
\begin{equation}
    \vec{x}^{0},  \, \vec{z}^{0} \, = \, \mathcal{A}^{*} \left( \tilde{\vec{y}} \right),
    \label{eq:x0_sense}
\end{equation}
although different task-specific initializations can be used.

\subsubsection{Lagrange Multipliers Initialization}
\label{sec:subsubsec2.2.2}
In our framework, the initialization of the Lagrange multiplier $\vec{u}^{0}$ involves the use of a neural network $\mathcal{G}_{\boldsymbol{\psi}}$ with  trainable parameters $\boldsymbol{\psi}$:
\begin{equation}
    \vec{u}^{0} = \mathcal{G}_{\boldsymbol{\psi}} \left( \vec{x}^{0} \right).
\end{equation}
The purpose of this network is to provide an initial estimation of the Lagrange multipliers based on the initial image guess $\vec{x}^{0}$. The architecture of the Lagrange multiplier initializer model draws inspiration from previous works, specifically the replication padding (ReplicationPadding) module \cite{Liu2022} and dilated convolutions \cite{yu2015multi}. The model consists of four layers of replication padding followed by a two-dimensional dilated convolution with a specified dilation factor and filter size. The output of this convolutional layer is then passed through a series of two-dimensional $1 \times 1$ convolutions, followed by a rectified linear unit (ReLU) activation function. 

By incorporating these initialization steps, vSHARP sets the initial values of $\vec{x}$, $\vec{z}$, and $\vec{u}$, enabling the iterative reconstruction process to begin from a reasonable starting point.

\subsubsection{Initialization of Optimization Parameters}

Instead of making the penalty $\boldsymbol{\rho}$ a hyperparameter, its chosen as a trainable parameter of vSHARP and it is initialized from the truncated normal distribution $\mathcal{N}_{\text{trunc}}(0, \,  \mat{I}_{T})$. Furthermore, vSHARP also learns the values of the step sizes $\boldsymbol{\eta}$ for the DCGD step, which is also initialized from $\mathcal{N}_{\text{trunc}}(0, \, \mat{I}_{T_x})$.

\subsection{End-to-end Pipeline of vSHARP}
\label{sec:subsec3.3}

\begin{algorithm*}
\caption{vSHARP: variable Splitting Half-quadratic ADMM algorithm for Reconstruction of inverse-Problems}
\label{alg:vSHARP}

\textbf{Input:} 
\newline \medskip  $T \, > \, 0,\, T_{\vec{x}} \, > \,0$, $\boldsymbol{{\rho}} \in \mathbb{R}^{T}$, $\boldsymbol{\eta} \in \mathbb{C}^{T_{\vec{x}}}$ ;


\textbf{Initialize:} 
%
$\vec{x}^{0} \, = \, \mathcal{A}^{*}(\Tilde{\vec{y}})$ ; \newline
%
$\vec{z}^0 \, = \, \vec{x}^0$ ; \newline 
%
$\vec{u}^{0} \,  = \, \mathcal{G}_{\boldsymbol{\psi}}(\vec{x}^{0})$ ;  
%
 
\For{$t\, = \, 0$ \textup{to} $T-1 $}{
    Update $\vec{z}^{t+1}  \, = \, \mathcal{R}_{\theta_{t + 1}}( \vec{z}^{t},\, \vec{x}^{t},\,\frac{\vec{u}^t}{{\rho}_{t+1}})$ \tcp*{$\mathcal{R}_{\theta_{t}}$  denoiser at $t$-th step}
    

    Update $\vec{x}^{t+1} \, = \, \text{DCGC}_{({\rho}_{t+1}, T_x, \boldsymbol{\eta})}(\vec{z}^{t+1}, \, \vec{x}^{t}, \, \vec{u}_{t})$ \tcp*{Data Consistency via Gradient Descent}

    
    Update $\vec{u}^{t+1} = \vec{u}^t + {\rho}_{t+1} (\vec{x}^{t+1} - \vec{z}^{t+1})$ ;
}


\textbf{Output:} $ \left\{ \vec{x}^{t}\right\}_{t=1}^{T}$ ;
\end{algorithm*}

\begin{figure*}[!hbt]
\centering
\includegraphics[width=1\textwidth]{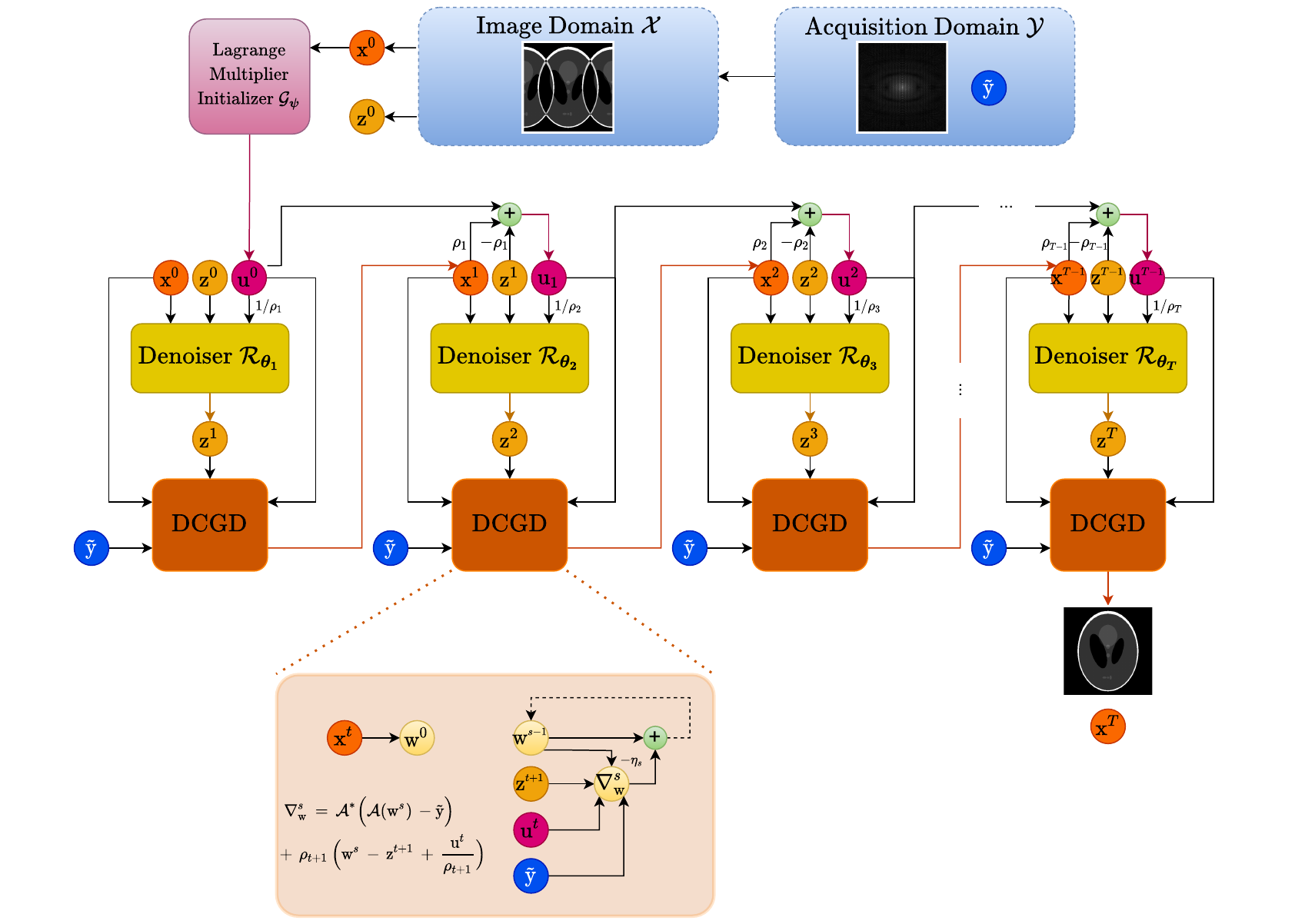}
\caption{Graphical Illustration of vSHARP. Utilizing acquired measurements $\Tilde{\vec{y}}$ from the acquisition domain $\mathcal{Y}$, initial estimations are made for the image $\vec{x}_0$ and the auxiliary variable $\vec{z}_0$ (in the image domain $\mathcal{X}$. The Lagrange Multiplier Initializer Network $\mathcal{G}_{\boldsymbol{\psi}}$ \cite{Yiasemis_2022_CVPR} generates an initialization $\vec{u}_0$ for the Lagrange multipliers, using $\vec{x}_0$ as input. These images, namely $\vec{z}_0$, $\vec{x}_0$, and $\vec{u}_0$, serve as the initial quantities in the unrolled ADMM algorithm spanning $T$ iterations. During each iteration, a DL-based denoiser $\mathcal{R}_{\boldsymbol{\theta}_t}$ (in this case, a U-Net) is employed to refine $\vec{z}_t$, while $\vec{x}_t$ undergoes optimization through a differentiable Data Consistency via Gradient Descent (DCGD) algorithm. The DCGD algorithm itself is further unrolled over $T_{\vec{x}}$ iterations. As the result of these operations, vSHARP produces $\vec{x}_T$ as the predicted reconstructed image.}
\label{fig:diagram}
\end{figure*}

The complete vSHARP reconstruction process follows a systematic pipeline, leveraging the strategies outlined in the previous sections. Algorithm \ref{alg:vSHARP}, combines all aforementioned strategies and outlines the end-to-end vSHARP pipeline. We also provide a graphical illustration of the end-to-end pipeline in \Figure{diagram}.

At each iteration, vSHARP refines the estimates of $\vec{x}$, $\vec{z}$, and $\vec{u}$, moving closer to an accurate reconstruction.  vSHARP outputs a sequence of reconstructions obtained at all optimization time-steps $ \left\{ \vec{x}^{t}\right\}_{t=1}^{T} $. Although $\vec{x}^{*}=\vec{x}_{T}$ is the prediction of the underlying image, all time-step predictions can be used for the loss computation in \Section{subsec3.4} to train the model.

In our experimental setup (\Section{sec4}), we particularize vSHARP for two ill-posed inverse problems, specifically accelerated parallel (static and dynamic) MRI reconstruction.

\subsection{vSHARP Training}
\label{sec:subsec3.4}

Let $\vec{x}$ be the ground truth image, $\vec{y}$ and $\vec{y}^{T}$ the corresponding ground truth and predicted measurements (obtained from the predicted image
$\vec{x}^{T}$ using some relevant transformation for the specific task). Then, for arbitrary loss functions $\mathcal{L}_{\mathcal{X}}$ and $\mathcal{L}_{\mathcal{Y}}$, computed in the image and measurement domains respectively, vSHARP is trained by computing the following:
\begin{equation}
\begin{gathered}
     \mathcal{L}\left(\vec{x},  \vec{x}^{t}\right) = \sum_{t=1}^{T} w_{t}  \mathcal{L}_{\mathcal{X}}\left(\vec{x},  \vec{x}^{t}\right) + \mathcal{L}_{\mathcal{Y}}\left(\vec{y},  \vec{y}^{T}\right) \\  w_{t} = 10^{\frac{t-T}{T - 1}}, \quad  t=1,\cdots,T.
     \label{eq:loss}
\end{gathered}
\end{equation}

\section{Experiments}
\label{sec:sec4}

\subsection{Accelerated Parallel MRI Reconstruction}
\label{sec:subsec4.1}

\subsubsection{Defining the Problem}
\label{sec:subsubsec4.1.1}

In the context of accelerated parallel MRI reconstruction, the primary objective is to generate a reconstructed complex image $\vec{x} \in \mathcal{X} = \mathbb{C}^{n}$ from undersampled $k$-space data collected from multiple ($n_c$) coils, denoted as $\tilde{\vec{y}} \in \mathcal{Y} = \mathbb{C}^{n \times n_{c}}$. An essential aspect of parallel MRI involves calculating coil sensitivity maps $\mat{C} = (\mat{C}_{1}, \mat{C}_{2}, \ldots, \mat{C}_{n_c}) \in \mathbb{C}^{n \times n_{c}}$, which indicate the spatial sensitivity of each individual coil. Typically, these maps are pre-computed from the central region of the multi-coil $k$-space, often referred to as the autocalibration (ACS) region.

The forward operator $\mathcal{A}_{\mat{M},\mat{C}}:\mathbb{C}^{n} \rightarrow \mathbb{C}^{n \times n_c}$ and the adjoint operators $\mathcal{A}_{\mat{M},\mat{C}}^{*}:\mathbb{C}^{n \times n_c} \rightarrow \mathbb{C}^{n}$ for accelerated parallel MRI can be defined as follows:
\begin{equation}
    \mathcal{A}_{\mat{M},\mat{C}}(\vec{x}) = \mat{M} \circ \mathcal{F} \circ \mathcal{E}_{\mat{C}}(\vec{x}), \quad \mathcal{A}_{\mat{M},\mat{C}}^{*} = \mathcal{R}_{\mat{C}} \circ \mathcal{F}^{-1} \circ \mat{M}.
    \label{eq:operators}
\end{equation}
In the forward operator, an undersampling operator $\mat{M}$ is combined with the two-dimensional Fourier transform $\mathcal{F}$ and the coil-encoding operator $\mathcal{E}_{\mat{C}}:\mathbb{C}^{n} \rightarrow \mathbb{C}^{n \times n_c}$, which transforms the image $\vec{x}\in\mathbb{C}^{n}$ into separate coil images using $\mat{C}$ as shown below:
\begin{equation}
    \mathcal{E}_{\mat{C}}(\vec{x})\,=\,\big(\mat{C}_{1}\vec{x},\,\mat{C}_{2}\vec{x},\cdots,\,\mat{C}_{n_c}\vec{x}\big).
    \label{eq:expand_op}
\end{equation}
On the other hand, the adjoint operator performs undersampling on the multi-coil $k$-space data, followed by the two-dimensional inverse Fourier transform $\mathcal{F}^{-1}$, and then applies the SENSE operator $\mathcal{R}_{\mat{C}}:\mathbb{C}^{n \times n_c} \rightarrow \mathbb{C}^{n}$, which combines individual coil images $\vec{z}\in\mathbb{C}^{n \times n_c}$ using $\mat{C}$:
\begin{equation}
    \mathcal{R}_{\mat{C}}(\vec{z}) = \sum_{k=1}^{n_c} \mat{C}_{k}^{*} \vec{z}_{k}.
    \label{eq:reduce_op}
\end{equation}

\subsubsection{Estimation and Refinement of Sensitivity Maps}
\label{sec:subsubsec4.1.2}

In our experimental framework, we employ the ACS $k$-space data to initially compute $\tilde{\mat{C}}$ as an estimation of the sensitivity maps. Subsequently, we employ a deep learning-based model $\mathcal{S}_{\theta_{\mathcal{S}}}$ with trainable parameters $\theta_{\mathcal{S}}$ to refine the sensitivity maps, utilizing $\tilde{\mat{C}}$ as input:
\begin{equation}
    \mat{C}_{k} = \mathcal{S}_{\theta_{\mathcal{S}}} ( \tilde{\mat{C}}_{k} ), \quad k =  1, \cdots, n_c.
\end{equation}
\noindent
This module is trained end-to-end along with vSHARP (incrorporated before step 2 in Algorithm \ref{alg:vSHARP}).

\subsubsection{Experimental Setup}
\label{sec:subsubsec4.1.3}

\paragraph{Datasets}
\label{sec:para5.1.3.1}

In the context of accelerated static MRI reconstruction  we utilized two publicly accessible datasets: the Calgary-Campinas (CC) brain dataset released as part of the Multi-Coil MRI Reconstruction (MCMRI) Challenge \cite{Beauferris2022} and the fastMRI T2 prostate dataset \cite{tibrewala2023fastmri}. Both datasets consist of  2D axial, multi-coil k-space measurements. The CC dataset encompasses 67 T1-weighted fully-sampled acquisitions, which we divided into training (40 volumes, 7332 2D slices), validation (10 volumes, 1560 2D slices), and test (10 volumes, 1560 2D slices) subsets. The fastMRI T2 prostate dataset comprises fully-sampled data acquired from three averages with GRAPPA factor of 2, from 312 subjects, partitioned into training (218 volumes, 6602 2D slices), validation (48 volumes, 1462 2D slices), and test (46 volumes, 1399 2D slices) sets.

\paragraph{Undersampling}
\label{sec:para5.1.3.2}

In order to comprehensively assess our model's capabilities, we incorporated the use of distinct undersampling patterns in our experiments.

Given that the available data was fully-sampled, we retrospectively generated undersampling masks to simulate acceleration factors of 4, 8, and 16. It is noteworthy that for each acceleration factor, we retained an 8\%, 4\%, and 2\% center fraction of the fully-sampled data within the autocalibration region. During the training phase, we applied random undersampling to the data, whereas during the inference phase, we evaluated the model's performance by undersampling the data across all acceleration levels.

For the CC dataset, we employed Variable Density Poisson disk masks, which were provided along with the dataset. On the other hand, for the fastMRI prostate data, we opted for vertical rectilinear equispaced undersampling \cite{yiasemis2023retrospective}. This choice was influenced by the practicality of implementing equispaced undersampling, which closely aligns with standard MRI scanner procedures.

\paragraph{Training Details}
\label{sec:para5.1.3.3}

Given these conditions, the training of vSHARP was executed utilizing the retrospectively undersampled $k$-space measurements $\tilde{\vec{y}} = \mat{M} (\vec{y})$, where $\vec{y}$ represents the fully-sampled $k$-space measurements. A ground truth image $\vec{x}$ was obtained from applying the inverse Fourier transform on $\vec{y}$ followed by the root-sum-of-squares (RSS) operation: 
\begin{equation}
        \vec{x} = \text{RSS} \circ \mathcal{F}^{-1} (\vec{y}) = \left( \sum_{k=1}^{n_c} |\mathcal{F}^{-1}(\vec{y}_k)|^{2} \right)^{1/2} \in \mathbb{R}^{n}.
\end{equation}
\noindent
It is worth noting that the RSS method was chosen as it has been demonstrated to be the optimal unbiased estimator for the underlying ground truth image \cite{Larsson2003}.

Furthermore, we initialized \eqref{eq:x0_sense} as $\vec{x}_0 = \vec{z}_0 = \mathcal{A}_{\mat{M},\mat{C}}^{*} (\tilde{\vec{y}})$.  For the purpose of computing the loss and obtaining predicted $k$-space measurements from the model, we transformed the predicted image $\vec{x}^{T}$ to the $k$-space domain using the following transformation:
\begin{equation}
    \vec{y}^{T} = \mathcal{F} \circ \mathcal{E}_{\mat{C}} (\vec{x}^{T}). 
\end{equation}

We computed a combination of various loss functions with equal weighting computed in both the image and $k$-space domains. To be specific, we formulated the training losses as:
\begin{equation}
\begin{gathered}
        \mathcal{L}_{\mathcal{X}} := \mathcal{L}_{1} + \text{SSIMLoss} + \text{HFEN}_{1} + \text{HFEN}_{2} \\ \mathcal{L}_{\mathcal{Y}} := \text{NMSE} + \text{NMAE}.
\end{gathered}
\end{equation}
\noindent
For the aforementioned loss functions, we utilized the loss definitions as follows:

\paragraph{Image Domain Loss Functions Definitions}

For images $\vec{v}, \vec{w}  \in \mathbb{R}^{n}$:

\begin{enumerate}[label=\Alph*, leftmargin=*]
\item \textbf{Absolute Differences Loss}
\begin{equation}
    \mathcal{L}_1(\vec{v}, \vec{w})  = ||\vec{v} - \vec{w} ||_1
\end{equation}

\item \textbf{Structural Similarity Index Measure Loss}
\begin{subequations}
\begin{equation}
\begin{gathered}
        \text{SSIMLoss}(\vec{v}, \vec{w}) = 1 - \text{SSIM}(\vec{v}, \vec{w}), \\
        \text{SSIM}(\vec{v},\,\vec{w}) = \\
    \frac{1}{M}\sum_{i=1}^{M} 
    \frac{(2\mu_{\vec{v}_i}\mu_{\vec{w}_i} + 0.01)(2\sigma_{\vec{v}_i\vec{w}_i} + 0.03)}{({\mu^2_{\vec{v}_i}} +{\mu^2_{\vec{w}_i}} + 0.01)({\sigma^2_{\vec{v}_i}} + {\sigma^2_{\vec{w}_i}} + 0.03)},
\end{gathered}
\end{equation}
\begin{equation}
\begin{gathered}
    \mu_{\vec{v}_i} =  \text{Mean}(\vec{v}_i),\quad \mu_{\vec{w}_i} =  \text{Mean}(\vec{w}_i),\\ \sigma^2_{\vec{v}_i} =  \text{Var}(\vec{v}_i),\quad \sigma^2_{\vec{w}_i} =  \text{Var}(\vec{w}_i), \\ \sigma_{\vec{v}_i\vec{w}_i} = \text{Cov}(\vec{v}_i, \vec{w}_i),
\end{gathered}
\end{equation}
\label{eq:ssim}
\end{subequations}
   
where $\vec{v}_i,\,\vec{w}_i,\,i=1,\cdots,M$ are $7 \times 7$ image windows from $\vec{v}$ and $\vec{w}$, respectively. 

\item \textbf{High Frequency Error Norm Loss}
\begin{equation}
\begin{gathered}
    \text{{HFEN}}_1(\vec{v},\, \vec{w})\, = \, \frac{|| \text{{LoG}}(\vec{v}) - \text{{LoG}}(\vec{w}) ||_1}{||\text{{LoG}}(\vec{v})||_1}, \\
    \text{{HFEN}}_2(\vec{v},\, \vec{w})\, = \, \frac{|| \text{{LoG}}(\vec{v}) - \text{{LoG}}(\vec{w}) ||_2}{||\text{{LoG}}(\vec{v})||_2},
    \label{eq:hfen}
\end{gathered}
\end{equation}
where where $\text{LoG}$ denotes a $15\times 15$ kernel filter of the Laplacian of a Gaussian distribution with standard deviation of 2.5.

\end{enumerate}

\paragraph{\textit{k}-space Domain Loss Functions}

For $k$-space data $\vec{v}, \vec{w}  \in \mathbb{C}^{n\times n_c}$:

\begin{enumerate}[label=\Alph*, leftmargin=*]
\item \textbf{Normalized Mean Squared Error (NMSE)}
\begin{equation}
    \text{NMSE} (\vec{v},\, \vec{w})\,= \, \frac{||\vec{v}\,-\,\vec{w}||_2^2}{||\vec{v}||_2^2} 
    \label{eq:nmse}
\end{equation}
\item \textbf{Normalized Mean Average Error (NMAE)}
\begin{equation}
    \text{NMAE} (\vec{v},\, \vec{w})\,= \, \frac{||\vec{v}\,-\,\vec{w}||_1}{||\vec{v}||_1} 
    \label{eq:nmae}
\end{equation}
\end{enumerate}

\paragraph{Model Implementation}
\label{sec:para5.1.3.4}

For accelerated parallel MRI reconstruction, we designed a vSHARP model with $T = 12$ ADMM optimization steps and $T_{\vec{x}} = 10$ DCGD steps. Within the $\vec{z}$-step denoising process, we utilized two-dimensional U-Nets \cite{ronneberger2015unet} with four scales and initiated the first scale with 32 convolutional filters. For the sensitivity maps refinement, we integrated a U-Net with four scales and 16 filters in the first scale.

\subsection{Accelerated Parallel Dynamic MRI Reconstruction}
\label{sec:subsec4.2}

\subsubsection{Defining the Problem}
\label{sec:subsubsec4.2.1}

In the context of accelerated parallel dynamic MRI Reconstruction, the objective becomes to generate a reconstructed complex image sequence $\vec{x} \in \mathcal{X} = \mathbb{C}^{n \times n_f }$ from a sequence of undersampled multi-coil $k$-space data collected $\tilde{\vec{y}} \in \mathcal{Y} = \mathbb{C}^{n \times n_{c} \times n_f}$, where $n_f$ denotes the length of the sequence (number of time-frames).

The forward operator $\mathcal{A}_{\mat{M},\mat{C}}:\mathbb{C}^{n \times n_f} \rightarrow \mathbb{C}^{n \times n_c \times n_f}$ and the adjoint operators $\mathcal{A}_{\mat{M},\mat{C}}^{*}:\mathbb{C}^{n \times n_c \times n_f} \rightarrow \mathbb{C}^{n  \times n_f}$ for accelerated parallel dynamic MRI have the same definitions as those in \eqref{eq:operators}, but also account for the temporal component, by applying the operators $\mathcal{R}_{\mat{C}}$, $\mathcal{E}_{\mat{C}}$, $\mathcal{F}$, $\mathcal{F}^{-1}$, $\mat{M}$ time-frame-wise.

\subsubsection{Experimental Setup}
\label{sec:subsubsec4.2.3}

\paragraph{Datasets}
\label{sec:para5.2.3.1}

For our dynamic MRI reconstruction experiments we utilized a cine cardiac MRI dataset from CMRxRecon challenge 2023 \cite{cmrxrecon, cmrxrecondataset}. The dataset comprised 473 scans of 4D multi-coil ($n_c = 10$) fully-sampled $k$-space acquisitions, split into training (203 scans), validation (111 scans), and test sets (159 scans), corresponding to 1,364, 731 and 1,090, respectively, 2D sequences.

\paragraph{Undersampling}
\label{sec:para5.2.3.2}

In line with the CMRxRecon provided undersampling patterns, we retrospectively applied rectilinear equispaced masks to simulate acceleration factors of 4, 8, and 16, with corresponding center fractions of 8\%, 4\%, and 2\%, respectively. Also following the CMRxRecon protocol, the same mask was applied to all time frames of the time sequence.

\paragraph{Training Details}
\label{sec:para5.2.3.3}
The training setup was consistent with the static case as outlined in \Section{para5.1.3.3}, with the only difference lying in the use of an additional loss component in the image domain:
\begin{equation}
\begin{split}
        \mathcal{L}_{\mathcal{X}} :=& \mathcal{L}_{1} + \text{SSIMLoss}  + 2 \cdot \text{SSIM3DLoss} \\
        &+ \text{HFEN}_{1} + \text{HFEN}_{2},
\end{split}
\end{equation}
\noindent
where SSIM3DLoss := 1 - SSIM3D represents the structural similarity index measure loss for volumes, calculated by using $7 \times 7 \times 7$ voxels instead of $7 \times 7$ windows in \eqref{eq:ssim}. All other quantities were computed for each time-frame and averaged. 

\paragraph{Model Implementation}
\label{sec:para5.2.3.4}

For accelerated parallel dynamic MRI reconstruction, we developed a three-dimensional adaptation of the vSHARP model from \Section{para5.1.3.4} to accommodate the dynamic nature of the data (2D + time). This involved replacing the two-dimensional image denoisers with three-dimensional ones. The model specifically employs $T = 10$ optimization steps using three-dimensional U-Nets with four scales and 32 filters in the first scale for the \textbf{z}-step and $T_{\vec{x}} = 8$ DCGD iterations. Additionally, for the Lagrange Multipliers Initializer module, we extended the implementation presented in \Section{subsubsec2.2.2} to handle 3D data using three-dimensional replication padding and convolutional modules. The sensitivity map module used a two-dimensional U-Net with four scales and 16 filters in the first scale.

\subsection{Statistical Tests}

For each combination of dataset, acceleration factor, and metric, we performed statistical tests to determine if the average performance of the best model was significantly better than that of other models. Specifically, we calculated the performance differences between the best model and the others. The Shapiro-Wilk (SW) test \cite{SHAPIRO1965} was used to check if these differences followed a normal distribution. If the differences were normally distributed ($p_{\text{sw}} > \alpha$ from the SW test), we applied a paired t-test; otherwise, we used the Wilcoxon signed-rank test \cite{conover1999practical}. Statistical significance was assessed using the $p$-value of the employed statistic, with $p$-values less than $\alpha$ indicating a significant difference. To avoid crowding the paper, instances where the best model was not significantly better ($p > \alpha$) are marked with an asterisk ($\star$) in our reported results. We set $\alpha = 0.05$ as the significance level.

\section{Experimental Results}
\label{sec:sec5}
For our experiments, including the model implementations, datasets handling, and all Deep Learning utilities, we employed the DIRECT toolkit \cite{DIRECTTOOLKIT}. Code for vSHARP can be found in \url{https://github.com/NKI-AI/direct/tree/main/direct/nn/vsharp}.

\subsection{Accelerated Parallel MRI Reconstruction}
\label{sec:subsec5.1}

\subsubsection{Comparative Analysis}
\label{sec:subsubsec5.1.1}
We conducted a comparison involving our vSHARP implementation (as detailed in \Section{para5.1.3.4}), alongside two non-physics-based baseline and two state-of-the-art physics-based unrolled MRI reconstruction techniques:

\begin{enumerate}
\item A U-Net model with four scales, where the first scale comprised 64 convolutional filters.
\item A transformer-based architecture, specifically the Uformer model \cite{Wang_2022_CVPR}.
\item An End-to-End Variational Network (E2EVarNet) \cite{Sriram2020} with 12 cascades and U-Net regularizers, incorporating four scales and initializing the first scale with 64 convolutional filters.
\item A Recurrent Variational Network (RVarNet) \cite{Yiasemis_2022_CVPR}, winning solution to the MCMRI Challenge \cite{Beauferris2022}, adhering to the architectural specifics outlined in the original publication.
\end{enumerate}

These specific hyperparameters for 2, 3, and 4 were chosen to ensure computational efficiency during training while for model 1, we opted for a large  U-Net configuration. To maintain consistency, a sensitivity map refinement module was incorporated into all approaches using identical hyperparameters as the one integrated with vSHARP (as detailed in \Section{subsubsec4.1.2}).

\paragraph{Training Details}

In our experimental setup for accelerated parallel MRI reconstruction, the optimization process for our proposed model and the comparison models was carried out using the Adam optimizer. The optimizer's hyperparameters were set as follows: $\beta_1 = 0.9$, $\beta_2 = 0.999$, and $\epsilon = 1e-8$. The experiments were conducted on separate NVIDIA A6000 or A100 GPUs. All models were trained with a batch size of 2 two-dimensional slices over 100,000 to 120,000 iterations. 

To initiate training, a warm-up schedule was implemented. This schedule linearly increased the learning rate to its initial value of 0.002 over the course of 1000 warm-up iterations, applied to both datasets.

Subsequently, a learning rate decay strategy was employed during training. Specifically, for the Calgary-Campinas experiments, the learning rate was reduced by a factor of 0.2 every 20,000 training iterations. In contrast, for the fastMRI prostate experiments, the learning rate decay occurred every 30,000 iterations.

\paragraph{Quantitative Results}
\label{para:6.1.1.1}

\setlength{\tabcolsep}{3pt}
\renewcommand{\arraystretch}{1.5}
\begin{table*}[!htb]
\centering
\caption{Average evaluation metrics on the Calgary-Campinas test set. An asterisk ($\star$) indicates that the average best method (highlighted in \textbf{bold}) did not show a statistically significant improvement over the corresponding method.}
\label{tab:metrics-cc}
\resizebox{\textwidth}{!}{%
\begin{tabular}{cccccccccc}
\hline
\multirow{3}{*}{\textbf{Model}} & \multicolumn{9}{c}{\textbf{Metrics}}                                                                                                                                                                                                                                          \\ \cline{2-10} 
                                & \multicolumn{3}{c}{\textbf{$4\times$}}                                                    & \multicolumn{3}{c}{\textbf{$8\times$}}                                                  & \multicolumn{3}{c}{\textbf{$16\times$}}                                                 \\
                                & SSIM                         & pSNR                       & NMSE                          & SSIM                         & pSNR                      & NMSE                         & SSIM                         & pSNR                      & NMSE                         \\ \hline
U-Net                           & 0.9413 $\pm$ 0.0091          & 34.63 $\pm$ 1.14           & 0.0086 $\pm$ 0.0018           & 0.9116 $\pm$ 0.0144          & 32.37 $\pm$ 1.07          & 0.0137 $\pm$ 0.0033          & 0.8599 $\pm$ 0.0167          & 29.53 $\pm$ 0.89          & 0.0247 $\pm$ 0.0053          \\
Uformer                         & 0.9401 $\pm$ 0.0098          & 35.12 $\pm$ 1.15           & 0.0079 $\pm$ 0.0019           & 0.9064 $\pm$ 0.0144          & 32.59 $\pm$ 1.13          & 0.0141 $\pm$ 0.0033          & 0.8574 $\pm$ 0.0207          & 30.15 $\pm$ 1.08          & 0.0248 $\pm$ 0.0057          \\
E2EVarNet                          & 0.9565 $\pm$ 0.0071          & 36.88 $\pm$ 1.19           & 0.0050 $\pm$ 0.0006           & 0.9345 $\pm$ 0.0086          & 34.42 $\pm$ 0.79          & 0.0087 $\pm$ 0.0012          & 0.8986 $\pm$ 0.0127          & 31.80 $\pm$ 0.86          & 0.0156 $\pm$ 0.0020          \\
RVarNet                         & \textbf{0.9641 $\pm$ 0.0056} & \textbf{37.82 $\pm$ 1.00}  & \textbf{0.0041 $\pm$ 0.0008}  & 0.9443 $\pm$ 0.0104          & 35.34 $\pm$ 1.21          & 0.0073 $\pm$ 0.0019          & 0.9114 $\pm$ 0.0136          & 32.61 $\pm$ 0.98          & 0.0132 $\pm$ 0.0031          \\
vSHARP                          & 0.9631 $\pm$ 0.0062          & 37.71 $\pm$ 0.94$^{\star}$ & 0.0043 $\pm$ 0.0009$^{\star}$ & \textbf{0.9491 $\pm$ 0.0107} & \textbf{35.84 $\pm$ 1.13} & \textbf{0.0066 $\pm$ 0.0018} & \textbf{0.9255 $\pm$ 0.0161} & \textbf{33.52 $\pm$ 1.20} & \textbf{0.0109 $\pm$ 0.0030} \\ \hline
\end{tabular}%
}
\end{table*}

For quantitative comparative evaluation, we adopted three well established metrics: the Structural Similarity Index Measure (SSIM), peak signal-to-noise ratio (pSNR), and normalized mean squared error (NMSE) as defined in \cite{yiasemis2023retrospective}.

In  \Figure{metrics_cc} and  \Figure{metrics_prostate} we report box plots with the evaluation results obtained from the test subsets from the Calgary Campinas and fastMRI prostate datasets across distinct acceleration factors (4x, 8x, and 16x) for all considered techniques. Additionally, Tables \ref{tab:metrics-cc} and \ref{tab:metrics-prostate} report the corresponding average metrics evaluation along with their respective standard deviations and statistical significance results.

Upon reviewing the results presented in \Figure{metrics_cc} and \Table{metrics-cc} for the CC dataset, it is apparent that the vSHARP technique demonstrates competitive performance. Particularly, it showcases high SSIM and  pSNR and low NMSE values for all acceleration factors. While RVarNet slightly outperforms vSHARP at 4x acceleration (not found to be significantly better in terms of pSNR and NMSE), vSHARP remains a robust contender with compelling results across all factors.

\renewcommand{\arraystretch}{1.5}
\begin{table*}[!htb]
\centering
\caption{Average evaluation metrics on the fastMRI T2 prostate test set. An asterisk ($\star$) indicates that the average best method (highlighted in \textbf{bold}) did not show a statistically significant improvement over the corresponding method.}
\label{tab:metrics-prostate}
\resizebox{\textwidth}{!}{%
\begin{tabular}{cccccccccc}
\hline
\multirow{3}{*}{\textbf{Model}} & \multicolumn{9}{c}{\textbf{Metrics}} \\ \cline{2-10} 
 & \multicolumn{3}{c}{\textbf{$4\times$}} & \multicolumn{3}{c}{\textbf{$8\times$}} & \multicolumn{3}{c}{\textbf{$16\times$}} \\
 & SSIM & pSNR & NMSE & SSIM & pSNR & NMSE & SSIM & pSNR & NMSE \\ \hline
U-Net & 0.8567 $\pm$ 0.0317 & 32.07 $\pm$ 1.85 & 0.0165 $\pm$ 0.0061 & 0.7694 $\pm$ 0.0433 & 29.38 $\pm$ 1.62 & 0.0299 $\pm$ 0.0077 & 0.6719 $\pm$ 0.0501 & 26.55 $\pm$ 1.55 & 0.0567 $\pm$ 0.0122 \\
Uformer & 0.8367 $\pm$ 0.0379 & 28.57 $\pm$ 2.01 & 0.0365 $\pm$ 0.0131 & 0.7327 $\pm$ 0.0451 & 26.84 $\pm$ 1.78 & 0.0535 $\pm$ 0.0144 & 0.6396 $\pm$ 0.0507 & 25.13 $\pm$ 1.62 & 0.0784 $\pm$ 0.0154 \\
E2EVarNet & 0.9099 $\pm$ 0.0247 & 35.94 $\pm$ 1.69$^\star$ & 0.0066 $\pm$ 0.0019 $^\star$ & 0.8318 $\pm$ 0.0369 & 32.13 $\pm$ 1.50 & 0.0158 $\pm$ 0.0039 & 0.7332 $\pm$ 0.0451 & 28.66 $\pm$ 1.39 & 0.0348 $\pm$ 0.0072 \\
RVarNet & 0.9013 $\pm$ 0.0249 & 35.39 $\pm$ 1.60 & 0.0075 $\pm$ 0.0020 & 0.8104 $\pm$ 0.0386 & 31.35 $\pm$ 1.51 & 0.0190 $\pm$ 0.0046 & 0.6943 $\pm$ 0.0461 & 27.57 $\pm$ 1.39 & 0.0447 $\pm$ 0.0089 \\
vSHARP & \textbf{0.9141 $\pm$ 0.0245} & \textbf{35.95 $\pm$ 1.65} & \textbf{0.0066 $\pm$ 0.0018} & \textbf{0.8457 $\pm$ 0.0371} & \textbf{32.77 $\pm$ 1.57} & \textbf{0.0138 $\pm$ 0.0038} & \textbf{0.7565 $\pm$ 0.0461} & \textbf{29.44 $\pm$ 1.46} & \textbf{0.0293 $\pm$ 0.0071} \\ \hline
\end{tabular}%
}
\end{table*}

Turning to \Figure{metrics_prostate} and \Table{metrics-prostate}, which showcase the quantitative results for the fastMRI prostate dataset, vSHARP once again emerges as a top performer. It achieves the highest SSIM and pSNR values, along with the lowest NMSE values, with statistical significance across all acceleration factors, except for the pSNR and NMSE values of E2EVarnet at $R=4$, which were not significant.

\paragraph{Qualitative Results}

\begin{figure*}[!hbt]
    \centering
    \includegraphics[width=1\textwidth]{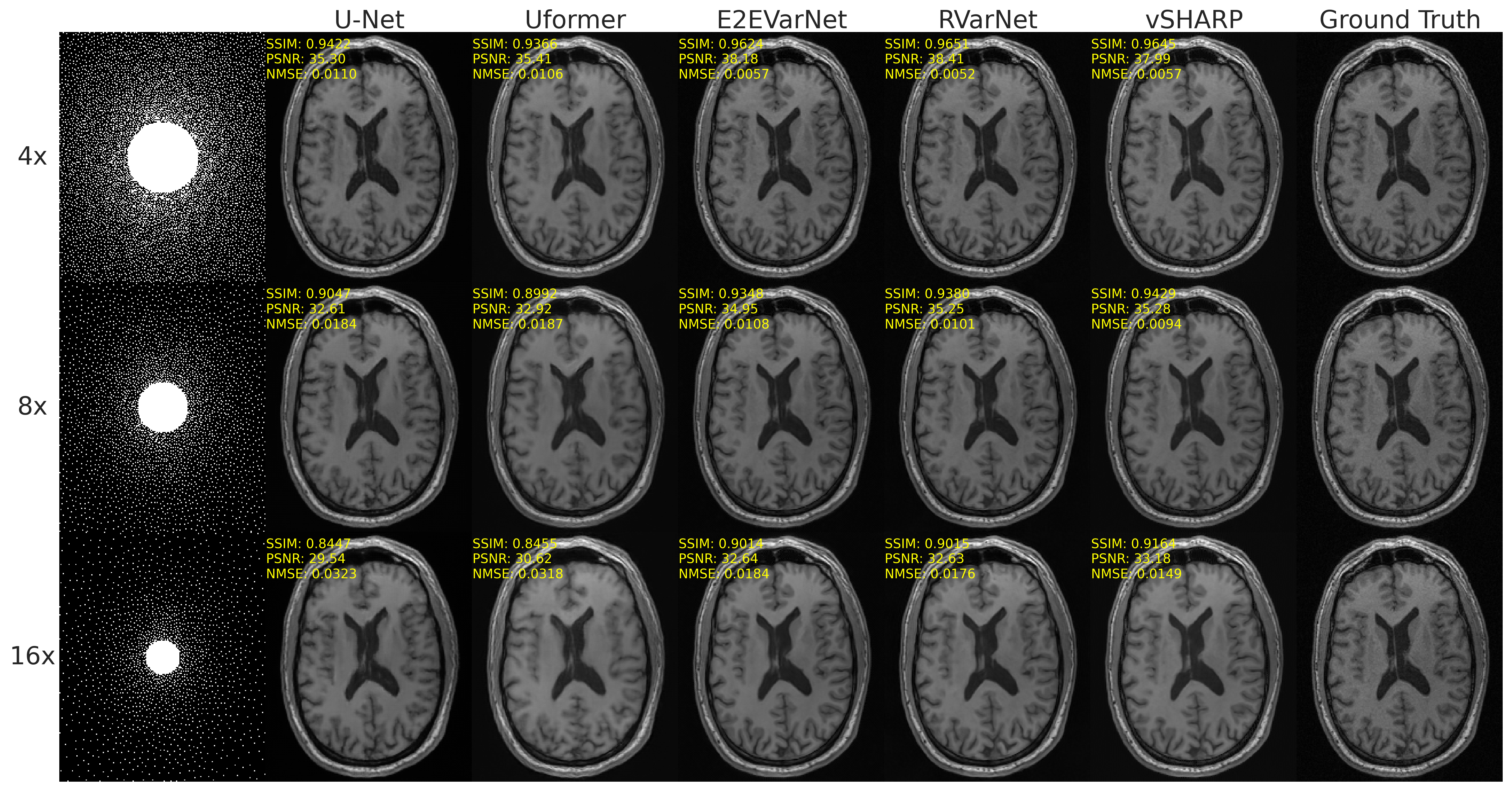}
    \caption{Sample reconstructions from the Calgary-Campinas test set.}
    \label{fig:recons_cc}
\end{figure*}

\begin{figure*}[!htb]
    \centering
    \includegraphics[width=1\textwidth]{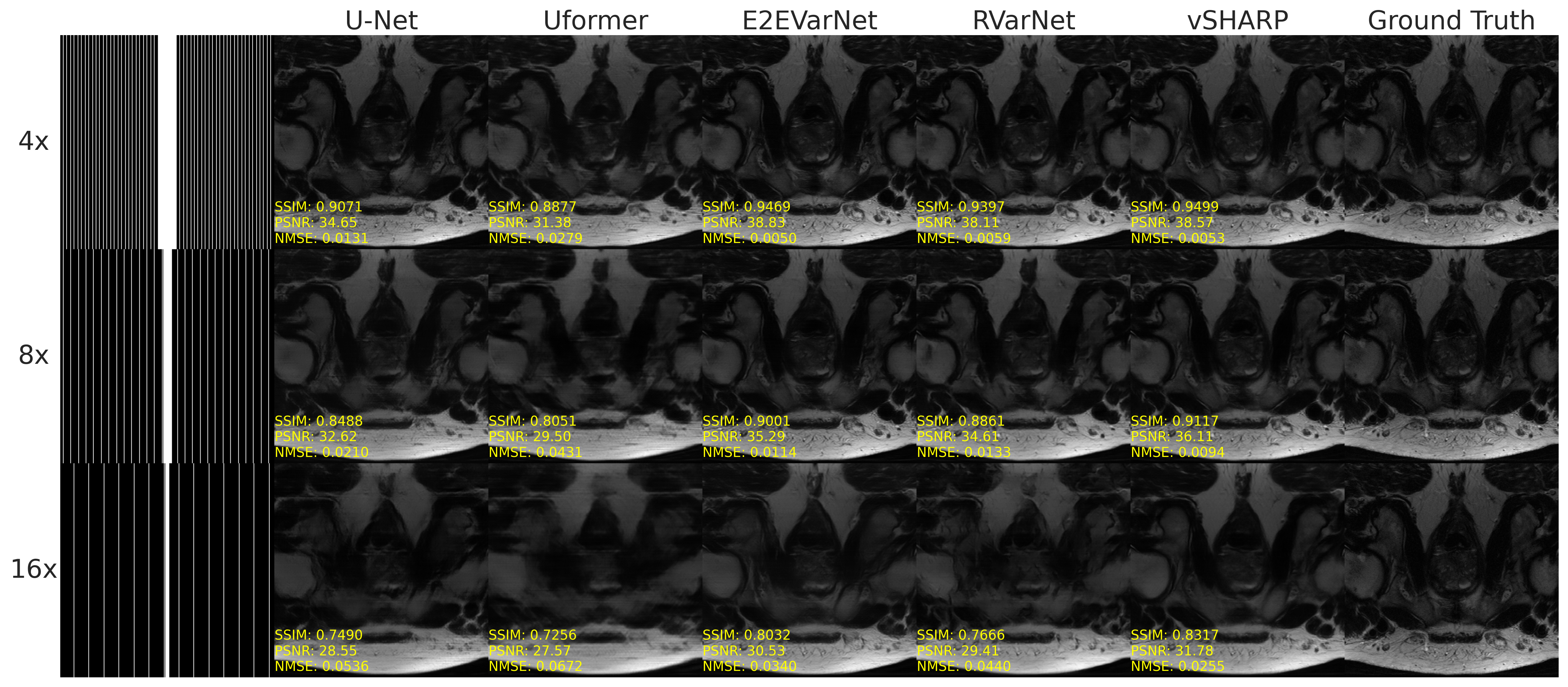}
    \caption{Sample reconstructions from the fastMRI prostate test set.}
    \label{fig:recons_prostate}
\end{figure*}


In Figures \ref{fig:recons_cc} and \ref{fig:recons_prostate}, we present sample reconstructions obtained from our experiments conducted on the CC and fastMRI prostate test sets, respectively, accompanied by their corresponding undersampling patterns for all acceleration factors, and ground truth reconstructions. Both Figures illustrate that vSHARP consistently delivers reconstructions with superior visual fidelity, underscoring its effectiveness, even for high acceleration factors such as 16$\times$.

\subsection{Accelerated Parallel Dynamic MRI Reconstruction}
\label{sec:subsec5.2}

\subsubsection{Comparative Analysis}
\label{sec:subsubsec5.2.1}
To assess the three-dimensional adaptation of vSHARP, using the parameters detailed in \Section{para5.2.3.4}, we compared against the following:

\begin{enumerate}
\item A three-dimensional U-Net with four scales (64 convolutional filters in the first scale).
\item An E2EVarNet \cite{Sriram2020} with 12 cascades, extended to 3D by employing three-dimensional U-Nets with four scales and initializing the first scale with 64 convolutional filters, as regularizers.
\item A RVarNet \cite{Yiasemis_2022_CVPR}, extended to 3D, by replacing two-dimensional convolutional recurrent unis with three-dimensional ones. To facilitate GPU memory constraints, we deployed 12 recurrent steps, hidden states with 32 channels, and 3 recurrent layers. All other hyperparameters were consistent with the original publication \cite{Yiasemis_2022_CVPR}.
\item A Variable Splitting Network (VS-Net) which adapts Alternating Minimization \cite{alternate_min} to solve a denoising and a DC step similar to vSHARP steps. More specifically, for the denoising step we employed the same denoisers as for vSHARP, and for the DC step employed a gradient descent scheme, same as vSHARP. The detailed architecture can be found in \url{https://github.com/NKI-AI/direct/tree/main/direct/nn/varsplitnet}.
\end{enumerate}
\noindent
For sensitivity map refinement we used an identical model to our 3D vSHARP (see \Section{para5.2.3.4}). 

\paragraph{Training Details}

For the optimization process in this section we used an Adam optimizer with  $\beta_1 = 0.9$, $\beta_2 = 0.999$, and $\epsilon = 10^{-8}$. The experiments were conducted on separate three NVIDIA A100 40GB GPUs with a batch size of one two-dimensional sequence over 200,000 iterations. An identical warm-up schedule to the static experiments (in \Section{subsec5.1}) was employed. The learning rate was decayed  by a factor of 0.8 every 30,000 training iterations.

\paragraph{Quantitative Results}

To quantitatively evaluate the performance in the dynamic reconstruction setting, we calculated SSIM, pSNR, and NMSE values by comparing the ground truths with the predictions. These metrics were computed for each time-frame in the sequence and then averaged. The evaluation metrics are presented in \Figure{metrics_cardiac}, while \Table{metrics-cardiac} summarizes the average evaluation metrics obtained from the test set. According to \Figure{metrics_cardiac} and \Table{metrics-cardiac}, the vSHARP 3D method consistently outperformed the other 3D models at all acceleration factors, reflected in the statistically significant higher SSIM and pSNR values, as well as the lower NMSE values. 

\paragraph{Qualitative Results}

\begin{figure*}[!htb]
    \centering
    \includegraphics[width=0.95\textwidth]{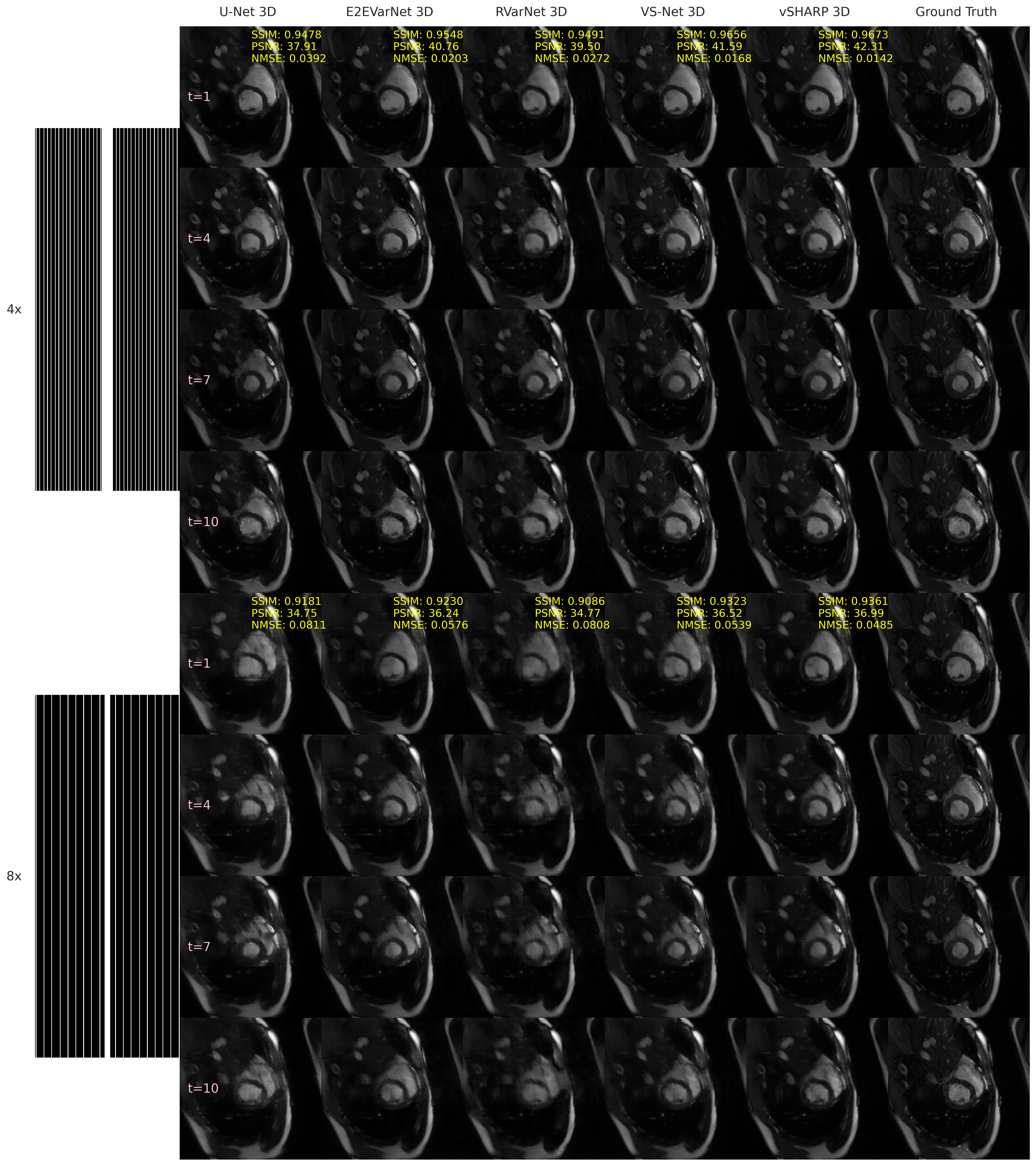}
    \caption{Sample reconstructions (cropped region of interest) from the CMRxRecon cardiac cine test set in the dynamic MRI reconstruction experiments. Only time-steps $t=1, 4, 7, 10$ (out of 12) are shown.}
    \label{fig:recons_cardiac}
\end{figure*}

\Figure{recons_cardiac} shows sample reconstructions from our experiments on the CMRxRecon cardiac cine test set at 4x and 8x accelerations of a single time sequence of a slice. Visual inspection indicates that vSHARP 3D provides superior image quality, retaining finer details and exhibiting fewer artifacts compared to the other methods. These qualitative improvements align with the quantitative metrics, underscoring the robustness and effectiveness of vSHARP  for accelerated dynamic MRI reconstruction tasks.

\renewcommand{\arraystretch}{1.5}
\begin{table*}[!htb]
\centering
\caption{Average evaluation on the CMRxRecon cardiac cine test set in the accelerated dynamic MRI reconstruction context. An asterisk ($\star$) indicates that the average best method (highlighted in \textbf{bold}) did not show a statistically significant improvement over the corresponding method.}
\label{tab:metrics-cardiac}
\resizebox{\textwidth}{!}{%
\begin{tabular}{cccccccccc}
\hline
\multirow{3}{*}{\textbf{Model}} & \multicolumn{9}{c}{\textbf{Metrics}} \\ \cline{2-10} 
 & \multicolumn{3}{c}{\textbf{$4\times$}} & \multicolumn{3}{c}{\textbf{$8\times$}} & \multicolumn{3}{c}{\textbf{$16\times$}} \\
 & SSIM & pSNR & NMSE & SSIM & pSNR & NMSE & SSIM & pSNR & NMSE \\ \hline
U-Net 3D & 0.9362 $\pm$ 0.0136 & 36.07 $\pm$ 1.65 & 0.0496 $\pm$ 0.0094 & 0.8914 $\pm$ 0.0200 & 32.97 $\pm$ 1.67 & 0.1011 $\pm$ 0.0187 & 0.8129 $\pm$ 0.0371 & 28.88 $\pm$ 2.10 & 0.2642 $\pm$ 0.0717 \\
VarNet 3D & 0.9384 $\pm$ 0.0127 & 38.63 $\pm$ 1.69 & 0.0280 $\pm$ 0.0076 & 0.8954 $\pm$ 0.0195 & 34.32 $\pm$ 1.72 & 0.0746 $\pm$ 0.0176 & 0.8280 $\pm$ 0.0340 & 29.98 $\pm$ 1.99 & 0.2044 $\pm$ 0.0530 \\
RVarNet 3D & 0.9298 $\pm$ 0.0150 & 37.53 $\pm$ 1.73 & 0.0362 $\pm$ 0.0109 & 0.8766 $\pm$ 0.0216 & 32.69 $\pm$ 1.70 & 0.1091 $\pm$ 0.0262 & 0.7940 $\pm$ 0.0366 & 28.06 $\pm$ 1.95 & 0.3155 $\pm$ 0.0693 \\
VS-Net 3D & 0.9573 $\pm$ 0.0101 & 39.75 $\pm$ 1.73 & 0.0216 $\pm$ 0.0061 & 0.9124 $\pm$ 0.0187 & 34.74 $\pm$ 1.73 & 0.0679 $\pm$ 0.0167 & 0.8453 $\pm$ 0.0344 & 30.19 $\pm$ 2.04 & 0.1962 $\pm$ 0.0564 \\
vSHARP 3D & \textbf{0.9627 $\pm$ 0.0095} & \textbf{40.45 $\pm$ 1.80} & \textbf{0.0185 $\pm$ 0.0057} & \textbf{0.9236 $\pm$ 0.0171} & \textbf{35.60 $\pm$ 1.77} & \textbf{0.0559 $\pm$ 0.0147} & \textbf{0.8572 $\pm$ 0.0322} & \textbf{30.65 $\pm$ 2.07} & \textbf{0.1775 $\pm$ 0.0560} \\ \hline
\end{tabular}%
}
\end{table*}

\subsection{Ablation Study}
\label{sec:subsec5.3}

\renewcommand{\arraystretch}{1.5}
\begin{table*}[!htb]
\centering
\caption{Average evaluation metrics for the ablation studies on the fastMRI T2 prostate test set. "Conv", "ResNet", and "Uformer" indicate the respective denoiser replacements for U-Net. "Small" refers to the vSHARP model with U-Net denoiser and modified hyperparameters ($T=8$ and $T_{\vec{x}}=6$).  An asterisk ($\star$) indicates that the average best method (highlighted in \textbf{bold}) did not show a statistically significant improvement over the corresponding method.}
\label{tab:ablation_metrics}
\resizebox{\textwidth}{!}{%
\begin{tabular}{cccccccccc}
\hline
\multirow{3}{*}{\textbf{\begin{tabular}[c]{@{}c@{}}Model\end{tabular}}} & \multicolumn{9}{c}{\textbf{Metrics}} \\ \cline{2-10} 
 & \multicolumn{3}{c}{\textbf{$4\times$}} & \multicolumn{3}{c}{\textbf{$8\times$}} & \multicolumn{3}{c}{\textbf{$16\times$}} \\
 & SSIM & pSNR & NMSE & SSIM & pSNR & NMSE & SSIM & pSNR & NMSE \\ \hline
Conv & 0.9125 $\pm$ 0.0242 & 35.37 $\pm$ 1.69 & 0.0075 $\pm$ 0.0020 & 0.8362 $\pm$ 0.0355 & 32.25 $\pm$ 1.51 & 0.0154 $\pm$ 0.0039 & 0.7273 $\pm$ 0.0475 & 28.49 $\pm$ 1.49 & 0.0364 $\pm$ 0.0083 \\
ResNet & 0.9146 $\pm$ 0.0241 & 35.83 $\pm$ 1.66 & 0.0068 $\pm$ 0.0018 & 0.8402 $\pm$ 0.0352 & 32.35 $\pm$ 1.51 & 0.0150 $\pm$ 0.0038 & 0.7329 $\pm$ 0.0473 & 28.60 $\pm$ 1.49 & 0.0353 $\pm$ 0.0076 \\
Uformer & 0.9147 $\pm$ 0.0250 & \textbf{36.23 $\pm$ 1.76} & \textbf{0.0062 $\pm$ 0.0019} & 0.8437 $\pm$ 0.0349$^\star$ & 32.64 $\pm$ 1.52$^\star$ & 0.0141 $\pm$ 0.0036$^\star$ & 0.7509 $\pm$ 0.0472 & 29.27 $\pm$ 1.52$^\star$ & 0.0304 $\pm$ 0.0070$^\star$ \\
Small & \textbf{0.9152 $\pm$ 0.0248} & 35.97 $\pm$ 1.69 & 0.0066 $\pm$ 0.0018 & 0.8423 $\pm$ 0.0358 & 32.56 $\pm$ 1.51 & 0.0143 $\pm$ 0.0036 & 0.7498 $\pm$ 0.0478 & 29.26 $\pm$ 1.52 & 0.0305 $\pm$ 0.0070$^\star$ \\
Original & 0.9141 $\pm$ 0.0245 & 35.95 $\pm$ 1.65 & 0.0066 $\pm$ 0.0018 & \textbf{0.8457 $\pm$ 0.0371} & \textbf{32.77 $\pm$ 1.57} & \textbf{0.0138 $\pm$ 0.0038} & \textbf{0.7565 $\pm$ 0.0461} & \textbf{29.44 $\pm$ 1.46} & \textbf{0.0293 $\pm$ 0.0071} \\ \hline
\end{tabular}%
}
\end{table*}

To further investigate vSHARP, we explored the following ablative settings in the 2D scenario for accelerated MRI reconstruction using the fastMRI prostate dataset. We tested different denoisers, keeping other hyperparameters identical to the original:
\begin{enumerate}
    \item vSHARP with a sequence of alternating convolutional and batch normalization layers, with PReLU activations after each batch normalization (except the last)  as denoisers.
    \item vSHARP with ResNet CNN structures \cite{Timofte_2017_CVPR_Workshops} as denoisers with  15 residual blocks with each having two 3x3 convolutional layers (128 hidden channels, ReLU and constant multiplication by 0.1 applied after the first layer).
    \item vSHARP with Uformers \cite{Wang_2022_CVPR} as denoisers with embedding dimension of 4 channels, encoder \{2, 2, 2\} depths.
\end{enumerate}
\noindent
Additionally, we considered a modification to the original hyperparameter choice:
\begin{enumerate}\addtocounter{enumi}{3}
\item A smaller vSHARP (with U-Nets as denoisers) using $T=8$ and $T_{\vec{x}}=6$.
\end{enumerate}

The results of our ablative studies experiments are illustrated in the form of box plots in \Figure{metrics_ablation} and the average metrics presented in \Table{ablation_metrics} along with statistical test results. In addition, in \Table{additional-info} are presented the average inference times as well as model parameter count for each considered ablative method.

\renewcommand{\arraystretch}{2.}
\begin{table*}[!ht]
    \centering
    \caption{Comparison of average inference times and total number of parameters. All inference times were obtained using the same machine (A100 Nvidia GPU).}
    \label{tab:additional-info}
    \resizebox{1\textwidth}{!}{%
\begin{tabular}{cccccc}
\hline
\multicolumn{2}{c}{\textbf{Model}} & \textbf{\begin{tabular}[c]{@{}c@{}}Average Inference Time\\ per volume (s)\\ (CC brain dataset)\end{tabular}} & \textbf{\begin{tabular}[c]{@{}c@{}}Average Inference Time \\ per volume (s)\\ (fastMRI prostate dataset)\end{tabular}} & \textbf{\begin{tabular}[c]{@{}c@{}}Average Inference Time\\ per 4D (3D+time) volume (s)\\ (CMRxRecon cardiac dataset)\end{tabular}} & \textbf{\begin{tabular}[c]{@{}c@{}}Number of\\ Parameters\\ (M)\end{tabular}} \\ \hline
 \parbox[t]{2.5cm}{\multirow{9}{*}{\rotatebox[origin=c]{90}{\begin{tabular}[c]{@{}c@{}}\textbf{Accelerated} \\ \textbf{(Static)} \\ \textbf{MRI Reconstruction}\end{tabular}}}} & U-Net & 17.2 & 8.5 & - & 31 \\
 & Uformer & 24.2 & 7.8 & - & 5 \\
 & VarNet & 26.5 & 12.4 & - & 373 \\
 & RVarNet & 25.1 & 19.8 & - & 7 \\
 & vSHARP (Original) & 40.1 & 20.1 & - & 93 \\
 & vSHARP (Conv) & - & 20.5 & - & 4 \\
 & vSHARP (ResNet) & - & 24.2 & - & 7 \\
 & vSHARP (Uformer) & - & 29.5 & - & 1.5 \\
 & vSHARP (Small) & - & 13.6 & - & 62 \\ \hline
\parbox[t]{2.5cm}{\multirow{5}{*}{\rotatebox[origin=c]{90}{\begin{tabular}[c]{@{}c@{}}\textbf{Accelerated}\\ \textbf{Dynamic} \\ \textbf{MRI Reconstruction}\end{tabular}}}} & U-Net 3D & - & - & 10.4 & 90 \\
 & VarNet 3D & - & - & 9.5 & 270 \\
 & RVarNet 3D & - & - & 11.4 & 4 \\
 & VS-Net 3D & - & - & 14.0 & 226 \\
 & vSHARP 3D & - & - & 11.6 & 226 \\ \hline
\end{tabular}%
    }
\end{table*}

Our ablation results in \Figure{metrics_ablation} and \Table{ablation_metrics} indicate that the original vSHARP ($T=12$ and $T_{\vec{x}}=10$, with U-Net denoisers) was the best overall performer for acceleration factors of 8 and 16. For a $4\times$ acceleration, the smaller variant ($T=8$ and $T_{\vec{x}}=6$, with U-Net denoisers) and the Uformers-based model were the top performers, for the SSIM and pSNR/NMSE metrics, respectively. Additionally, all vSHARPs at all ablative settings performed similarly for the $4\times$ and $8\times$ acceleration factors, but models with ResNet CNNs or convolutional block denoisers were outperformed by the other models at $R=16$.

Comparing \Table{metrics-prostate} and \Table{ablation_metrics}, it is evident that for all acceleration factors, the vSHARP variants surpassed the models (U-Net, Uformer, RVarNet, E2EVarNet) used in the comparison experiments.

Regarding inference times, the smaller vSHARP model was the fastest, processing in 13.6 seconds. Although the Uformer variant of the vSHARP model achieved competitive quantitative metrics, it had the longest inference time of 29.5 seconds, highlighting a trade-off between speed and performance.

Comparing the original vSHARP variant with the reduced hyperparameter version (Small), the latter offers slightly higher fidelity metrics at an acceleration factor of 4, while the original larger model remained consistently better at $R=8$ and 16. Nevertheless, the smaller variant reduced the inference time from 20.1 seconds (original vSHARP) to 13.6 seconds per prostate volume, making it ideal for scenarios requiring quick results with good performance.

For further assessment, we illustrate in \Figure{recons_prostate_ablation} sample reconstructions for a case from the test set.

\subsection{Additional Comparisons}

In \Table{additional-info}, we present supplementary comparisons beyond core performance metrics, focusing on inference time per volume in seconds (s) and the total number of parameters in millions (M) for various MRI reconstruction methods in both static and dynamic scenarios. All inference times were computed using the same A100 GPU for consistency.

Non-physics-based models, i.e. the U-Net (2D or 3D) and the Uformer demonstrated, notably lower inference times compared to physics-based unrolled models (vSHARP, RVarNet, E2EVarNet, VS-Net), despite not performing as well overall. Among the physics-based models, E2EVarNet had the lowest inference times. However, the differences were marginal when compared to the smaller vSHARP variant in the static scenario or to vSHARP 3D in the dynamic cases.

In the ablation study, alternative models with different denoisers (Conv, ResNet, Uformer) required more inference time than both the original vSHARP and its smaller variant with U-Nets. Specifically, vSHARP with Uformer or ResNet denoisers generally had the longest running times.

Regarding the number of parameters, E2EVarNet consistently required the most memory across both static and dynamic scenarios, with vSHARP being the second highest. In the static scenario, other methods had significantly fewer parameters.

The choice of the number of parameters for each method was carefully considered to ensure practical applicability during training. For instance, RVarNet’s recurrent nature demands substantial GPU memory during training, which limits the capacity to use a higher number of parameters.

\section{Conclusion and Discussion}
\label{sec:sec6}
In this work, we introduced vSHARP, a novel deep learning-based approach for solving ill-posed inverse problems in medical imaging, with a focus on accelerated MRI reconstruction. Our algorithm integrates the Half-Quadratic Variable Splitting method with the Alternating Direction Method of Multipliers, unrolling the optimization process and incorporating deep neural architectures for denoising and initialization.

We conducted a comprehensive evaluation of vSHARP against state-of-the-art methods for accelerated parallel static MRI reconstruction in \Section{subsec5.1}. Our results demonstrate that vSHARP achieves superior performance in terms of quantitative evaluation metrics (SSIM, pSNR, NMSE) on two public $k$-space datasets, the Calgary Campinas brain and fastMRI prostate datasets. vSHARP consistently outperformed existing methods, showcasing its effectiveness in addressing the complex problem of accelerated MRI reconstruction, particularly for high acceleration factors (8$\times$, 16$\times$).

Beyond static MRI reconstruction, we extended vSHARP to 3D to handle accelerated parallel dynamic MRI reconstruction, which involves reconstructing a sequence of images from undersampled multi-coil $k$-space data. This task is particularly challenging due to the temporal dimension. By adapting the forward and adjoint operators and the Lagrange multiplier module to account for the temporal component and using three-dimensional U-Nets for the z-step, we demonstrated that vSHARP can effectively reconstruct high-quality dynamic MRI sequences. Our experiments with the cine cardiac MRI dataset from the CMRxRecon 2023 challenge showed that vSHARP maintains its superior performance in dynamic contexts, with significant improvements in reconstruction quality over baseline methods as shown in \Section{subsec5.2}.

A noteworthy observation stemming from our results is the fact that RVarNet, despite its competitive performance against vSHARP on the CC dataset (\Section{subsec5.1}), performed significantly poorer on the fastMRI prostate (\Section{subsec5.1}) and CMRxRecon cardiac datasets (\Section{subsec5.2}). We attribute this discrepancy to its limited receptive field, which became a bottleneck when dealing with equispaced undersampling in the fastMRI prostate and CMRxRecon cardiac datasets, known for introducing substantial aliasing in the image domain. While increasing the number of parameters might enhance its performance, the recurrent nature of the network makes it challenging to fit in GPU memory during training. In contrast, both vSHARP (with U-Net denoisers) and E2EVarNet, which employed U-Nets, benefited from a larger receptive field, enabling them to handle the aliasing caused by the equispaced undersampling more effectively.

Despite vSHARP's superior performance in the main experiments, it was associated with longer reconstruction times compared to the baseline methods. However, one notable aspect of vSHARP is its adaptability. Our ablation study in \Section{subsec5.3}, focused on the prostate dataset, demonstrated that a "smaller" version of vSHARP with a reduced number of optimization blocks ($T=8$ instead of $T=12$) and fewer DCGD steps ($T_{\vec{x}}=6$ instead of $T_{\vec{x}}=10$) significantly reduced computational constraints such as memory usage (around 30\% less hyperparameters) and inference time (\Table{ablation_metrics}), while maintaining high reconstruction quality (\Figure{metrics_ablation}, \Table{ablation_metrics}). This variant not only enabled faster inference but also consistently outperformed the considered baselines (Tables \ref{tab:metrics-prostate} and \ref{tab:ablation_metrics}).

Furthermore, our ablation study in \Section{subsec5.3} explored various configurations of vSHARP to understand its adaptability. Specifically, we tested different denoiser choices (simple convolution blocks, ResNets, Uformers) while keeping other hyperparameters identical to the original. The results indicated that the vSHARP with Uformers model generally performed similarly to the vSHARPs with U-Net denoisers across all metrics and acceleration factors. Moreover, all ablative choices outperformed the considered baselines (\Figure{metrics_prostate}, \Table{metrics-prostate}).

Despite these promising results, it is important to acknowledge some limitations and considerations. First, vSHARP was not compared to every existing method for MRI reconstruction, which is infeasible due to the vast number of approaches. Nonetheless, we consider the examined methods representative of current baseline and state-of-the-art techniques for both static and dynamic MRI reconstruction. 

Moreover, vSHARP assumes knowledge of the forward and adjoint operators, which might not always be straightforward to compute. For instance, in non-Cartesian accelerated MRI reconstruction, the computation of the non-uniform fast Fourier transform and its adjoint can be computationally expensive due to the involved gridding process. Additionally, while vSHARP demonstrated competitive performance in MRI reconstruction, its applicability to other ill-posed inverse problems in medical imaging, such as sparse-view CT or CBCT reconstruction, should be explored in future research.

In conclusion, vSHARP represents a promising advancement in the field of medical imaging, particularly in accelerated MRI reconstruction. Its ability to combine the strengths of deep learning with mathematical optimization methods makes it a versatile and effective tool for solving ill-posed inverse problems, promising enhanced image quality and efficiency in medical imaging applications.

\section*{Funding Information}
This work was supported by institutional grants of the Dutch Cancer Society and of the Dutch Ministry of Health, Welfare and Sport.

\section*{Acknowledgments}
The authors would like to acknowledge the Research High Performance Computing (RHPC) facility of the Netherlands Cancer Institute (NKI).

\bibliographystyle{elsarticle-num} 
\bibliography{bibliography}

\newpage
\onecolumn

\renewcommand{\thetable}{A\arabic{table}\space}  
\renewcommand{\thefigure}{A\arabic{figure}\space}
\setcounter{figure}{0}
\setcounter{table}{0}

\appendix

\section{Additional Theory}
\label{sec:appendix1}
\subsection{ADMM Algorithm}
ADMM, in general, aims to solve problems possessing the following form:

\begin{equation}
    \min_{\vec{x} \in \mathcal{X}, \vec{z} \in \mathcal{Z}} f(\vec{x}) + g(\vec{z}), \quad \text{subject to } {B}(\vec{x}) + {C}(\vec{z}) = d, \quad d\in \mathcal{D},
    \label{eq:admm_general}
\end{equation}

where $\mathcal{X}$, $\mathcal{Z}$, $\mathcal{D}$ are Hilbert spaces, and ${B}: \mathcal{X} \rightarrow \mathcal{D}$, ${C}: \mathcal{Z} \rightarrow \mathcal{D}$ are operators on $\mathcal{X}$ and $\mathcal{Z}$, respectively.

The Lagrangian for \eqref{eq:admm_general} is formulated as follows:

\begin{equation}
    \mathcal{L}_{\rho}\left(\vec{x}, \vec{z}, \vec{u}\right) =  f(\vec{x}) + g(\vec{z}) + \left<\vec{u}, {B}(\vec{x}) + {C}(\vec{z}) - d \right>_{\mathcal{D}} + \frac{\rho}{2} \left|\left|{B}(\vec{x}) + {C}(\vec{z}) - d \right|\right|_2^2,
    \label{eq:admm_lagrangian_general}
\end{equation}

where $\rho > 0$, $\vec{u}$ are Lagrange multipliers, and $<\cdot, \cdot>_{\mathcal{D}}$ is an inner product on $\mathcal{D}$. In this setting, ADMM performs the following update over $T$ iterations:

\begin{subequations}
    \begin{equation}
        \vec{z}^{t+1}  = \argmin_{\vec{z} \in \mathcal{Z}} \mathcal{L}_{\rho} \left(\vec{x}^{t}, \vec{z}, \vec{u}^{t} \right)
        \label{eq:admm_general_z}
    \end{equation}
    \begin{equation}
        \vec{x}^{t+1}  = \argmin_{\vec{x} \in \mathcal{X}} \mathcal{L}_{\rho} \left(\vec{x}, \vec{z}^{t+1}, \vec{u}^{t} \right)
        \label{eq:admm_general_x}
    \end{equation}
    \begin{equation}
        \vec{u}^{t+1} = \vec{u}^t + \rho (\vec{x}^{t+1} - \vec{z}^{t+1}).
        \label{eq:admm_general_u}
    \end{equation}
\label{eq:admm_general_unrolled}
\end{subequations}

\subsection{ADMM for Inverse Problems}
A solution to an inverse problem as defined in \eqref{eq:forward_model} can be formulated as a solution to \eqref{eq:hqvs_problem}. By using the notations for inverse problems defined in \Section{sec2} and setting the following:

\begin{subequations}

    \begin{equation}
        \mathcal{X} = \mathcal{Z} = \mathcal{D}
    \end{equation}
    
    \begin{equation}
        f(\cdot) := \left|\left| \mathcal{A}(\cdot) - \tilde{\vec{y}}\right|\right|_2^2 + \lambda \mathcal{R}(\cdot), \quad g := 0
    \end{equation}

    \begin{equation}
        {B}  := \boldsymbol{1}_{\mathcal{X}}, \quad   {C}  := - \boldsymbol{1}_{\mathcal{X}}, \quad \vec{d} := 0,
    \end{equation}
\end{subequations}

we can derive the ADMM updates for solving inverse problems as stated in \eqref{eq:admm}.

\break
\section{Additional Figures}
\setcounter{figure}{0} 
\setcounter{table}{0} 
\renewcommand{\thefigure}{S\arabic{figure}}
\renewcommand{\thetable}{S\arabic{table}}

\label{sec:appendix2}

\begin{figure*}[!hbt]
    \centering
    \includegraphics[width=1\textwidth]{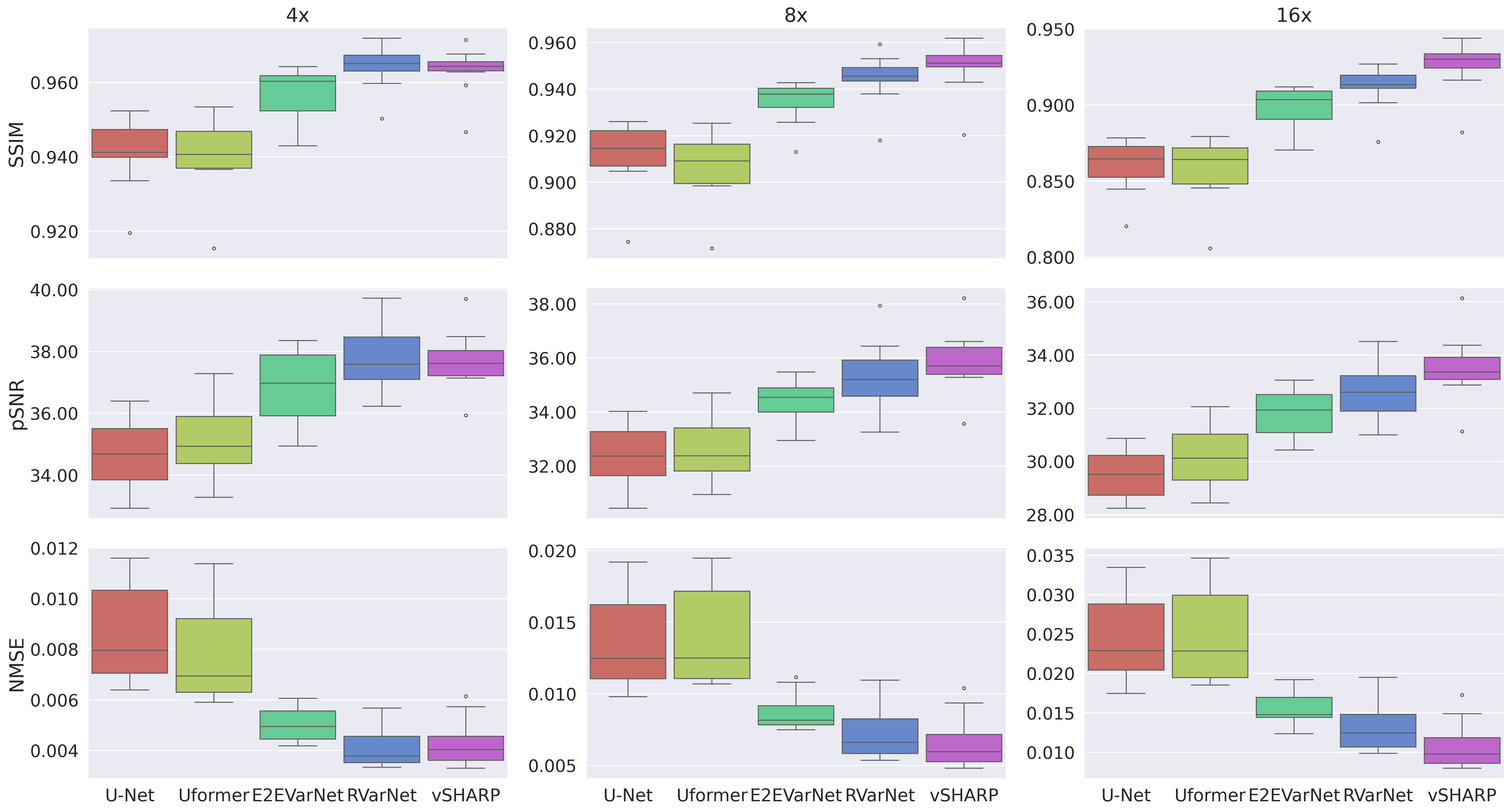}
    \caption{Evaluation metrics distribution on the Calgary-Campinas test set (box plots).}
    \label{fig:metrics_cc}
\end{figure*}

\begin{figure*}[!htb]
    \centering
    \includegraphics[width=1\textwidth]{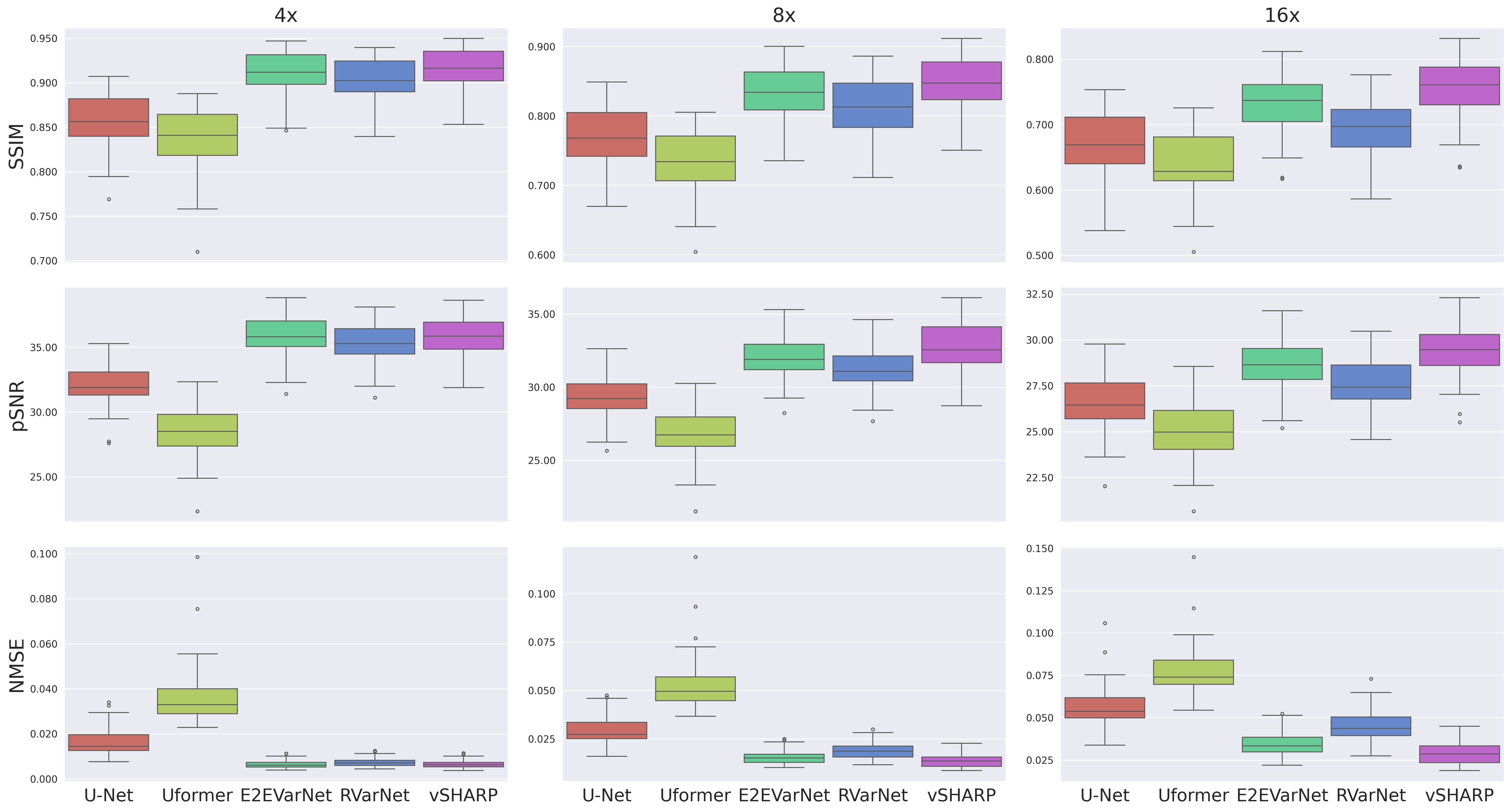}
    \caption{Evaluation metrics distribution on the fastMRI prostate T2 test set (box plots).}
    \label{fig:metrics_prostate}
\end{figure*}

\begin{figure*}[!htb]
    \centering
    \includegraphics[width=1\textwidth]{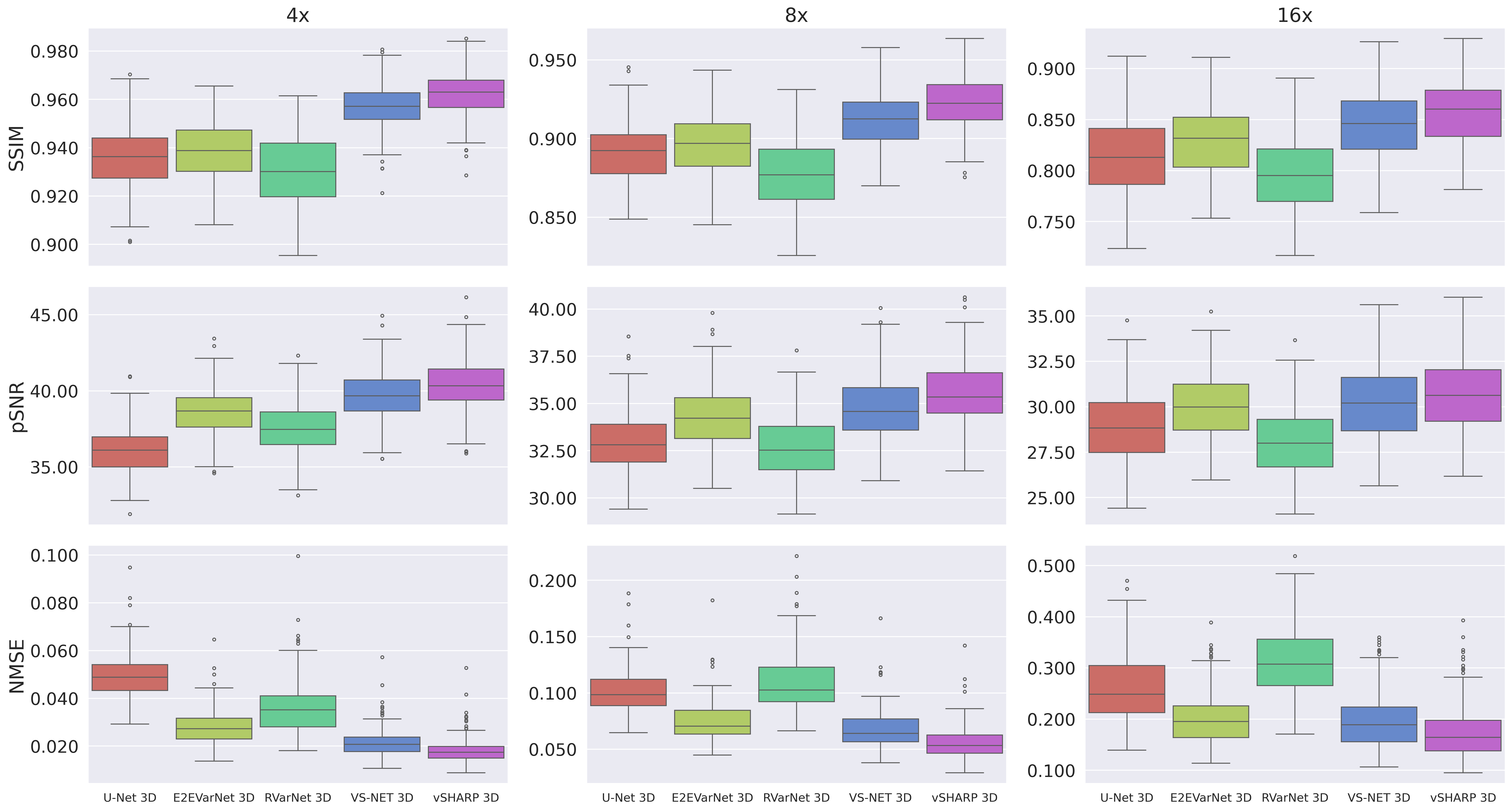}
    \caption{Evaluation metrics distribution on the CMRxRecon cardiac cine test set (box plots).}
    \label{fig:metrics_cardiac}
\end{figure*}

\begin{figure*}[!htb]
    \centering
    \includegraphics[width=1\textwidth]{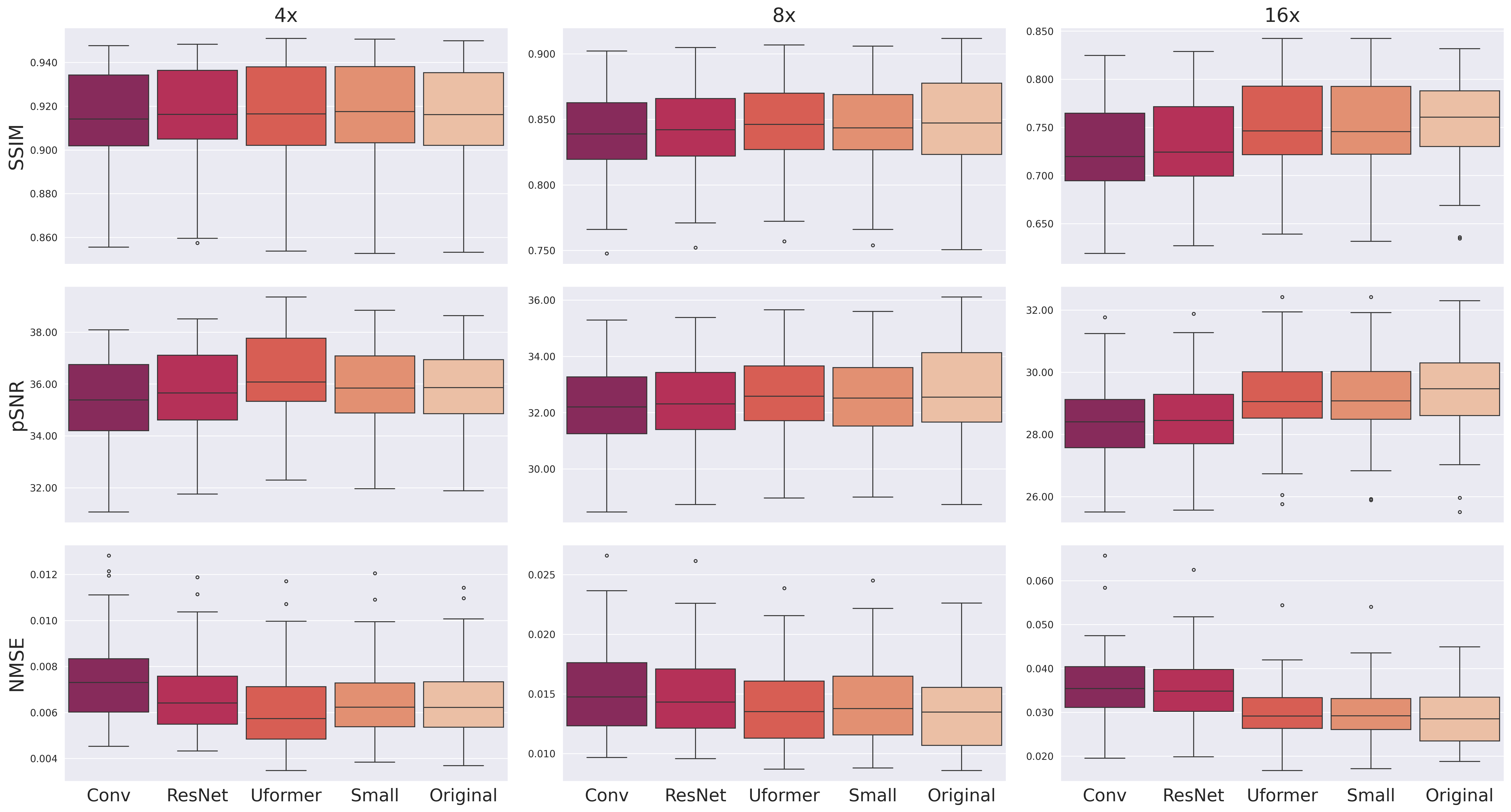}
    \caption{Evaluation metrics distribution on the fastMRI prostate T2 test set (box plots) for ablative settings.  "Conv", "ResNet", and "Uformer" indicate the respective denoiser replacements for U-Net. "Small" refers to the vSHARP model with U-Net denoiser and modified hyperparameters ($T=8$ and $T_{\vec{x}}=6$).}
    \label{fig:metrics_ablation}
\end{figure*}

\begin{figure*}[!htb]
    \centering
    \includegraphics[width=1\textwidth]{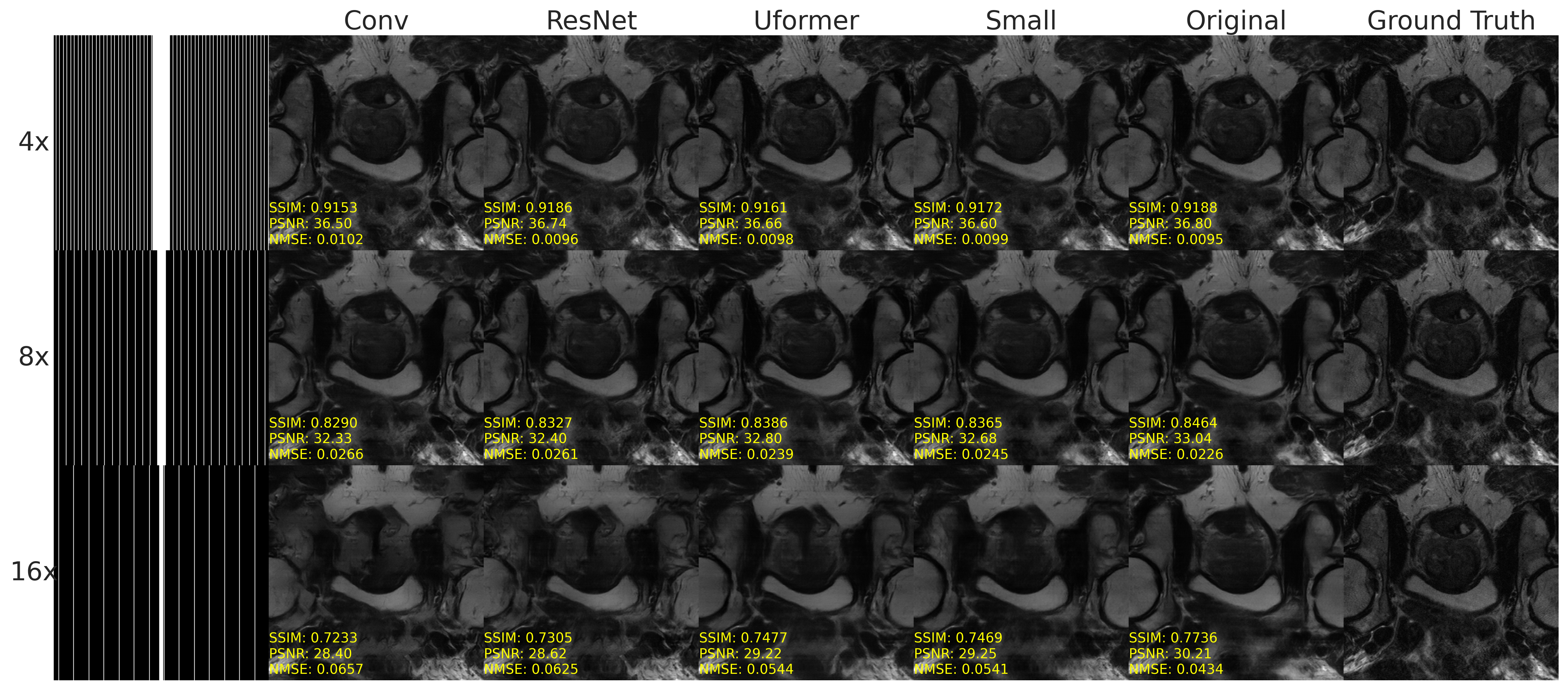}
    \caption{Sample reconstructions from the fastMRI prostate test set from the ablation studies.  "Conv", "ResNet", and "Uformer" indicate the respective denoiser replacements for U-Net. "Small" refers to the vSHARP model with U-Net denoiser and modified hyperparameters ($T=8$ and $T_{\vec{x}}=6$).}
    \label{fig:recons_prostate_ablation}
\end{figure*}

\end{document}